\documentclass[english,draftcls,onecolumn]{IEEEtran}

\usepackage{amsmath}
\usepackage{cite}
\usepackage{pb-diagram}
\usepackage{amssymb}
\usepackage{mathrsfs}
\usepackage{array}
\usepackage{enumerate}
\usepackage{color}
\usepackage{graphicx}
\usepackage{mathrsfs}
\usepackage{url}

\newcommand{\bv}{\mathbf{b}}
\newcommand{\gv}{\mathbf{g}}
\newcommand{\yv}{\mathbf{y}}

\newcommand{\QQ}{\mathbb{Q}}
\newcommand{\RR}{\mathbb{R}}

\newcommand{\Ac}{\mathcal{A}}

\newcommand{\Dc}{\mathcal{D}}

\newcommand{\<}{\langle}
\renewcommand{\>}{\rangle}						
\renewcommand{\mod}{\mbox{mod }}
\newcommand{\F}{\mathbb F}

\newcommand{\Q}{\mathbb{Q}}
\newcommand{\Z}{\mathbb{Z}}

\newcommand{\A}{\mathcal{A}}
\newcommand{\B}{\mathcal{B}}

\newcommand{\vf}{\varphi}

\newcommand{\qa}[3]{({#1},{#2})_{#3}}
\newcommand{\cc}[1]{{#1}^*}

\newcommand{\vc}[1]{\mathbf{#1}}

\newcommand{\bm}[4]{\begin{bmatrix}{#1} &{#2}\\{#3} &{#4}\end{bmatrix}}	

\newtheorem{example}{Example}
\newtheorem{lemma}{Lemma}
\newtheorem{claim}{Claim}
\newtheorem{corollary}{Corollary}
\newtheorem{defi}[lemma]{Definition}
\newtheorem{remark}[lemma]{Remark}

\begin{document}

\title{
Iterated Space-Time Code Constructions \\from Cyclic Algebras
\thanks{The authors are with the Division of Mathematical Sciences, School of Physical and Mathematical Sciences, Nanyang Technological University, Singapore.
Email: \{NMarkin,frederique\}@ntu.edu.sg.
Part of this work appeared in the proceedings of the 49th Allerton Conference on Communication, Control, and Computing, 2011, Illinois, USA, and in the proceedings of the International Conference on Signal Processing and Communications ({\em SPCOM 2012}), Bangalore, India.
}}

\author{Nadya Markin and Fr\'ed\'erique Oggier \\
}

\maketitle

\begin{abstract}
We propose a full-rate iterated space-time code construction, { {to design codes of $\Q$-rank $2n$ from cyclic algebra based codes of $\Q$-rank $n$}}. We give a condition for determining whether the resulting codes satisfy the full diversity property, and study their maximum likelihood decoding complexity with respect to sphere decoding. { { Particular emphasis is given to the asymmetric MIDO (multiple input double output) codes. }}In the process, we derive an interesting way of obtaining division algebras, and study their center and maximal subfield.
\begin{keywords}
Cyclic Algebras, Division Algebras, Fast Decodability, Full Diversity, Quaternion Algebras, Space-Time Coding.
\end{keywords}
\end{abstract}

%*************************************************************************************************************************%
%
% INTRO
%
%*************************************************************************************************************************%
\section{Introduction}
\label{intro}

We consider the problem arising in wireless communication of coding for a multiple antenna system, where the transmitter has twice more antennas than the 
receiver. Codes for that setting fall into the category of asymmetric space-time codes, and are usually formed of $2n\times 2n$ complex matrices.
The case when the receiver has exactly two antennas is called a MIDO channel (which stands for multiple input double output). Particular attention has 
recently been  paid to the MIDO case, due to its potential application to digital TV broadcast, where the end user carries a portable TV device.

Many good space-time codes are available in the literature, in that they satisfy the {\em full diversity} property requiring that $\det(X-X')\neq 0$ for any two matrices $X\neq X'$  in the codebook - ensuring good behavior at high signal-to-noise ratio (SNR), and the {\em  non-vanishing determinant (NVD)} property, which states that
$\min_{X\neq X'}|\det(X-X')|$ is bounded away from zero, even for an infinite space-time code, thus guaranteeing that the code performs well irrespectively of the alphabet size.

However, most of the codes enjoying the above properties in fact possess an underlying lattice structure, and might suffer from high maximum likelihood (ML) decoding complexity.
This triggered a new line of research focusing on designing codes with both good performance and reduced ML decoding  complexity via sphere decoding \cite{VB}, called {\em fast decodable codes} \cite{BHV}.
Analysis of conditions under which a code satisfies fast decodability can be found in e.g. \cite{JR,NR,LS}, where the notions
of {\emph{fast decodable}}, {\emph{group decodable}} and {\emph{conditionally group decodable}} are detailed. 
Several $4$-dimensional fast decodable MIDO codes have subsequently been studied, 
attempting to improve on the MIDO code in \cite{BHV}, which combines a quasi-orthogonal 
code with a twisted unitary transformation of another quasi-orthogonal code, and gets a 
lower decoding complexity, by sacrificing  the full diversity property, and consequently 
the stronger NVD property.
In \cite{SBM}, cyclic algebra based codewords are arranged in an Alamouti block code 
\cite{Alamouti} fashion to gain fast decodability for asymmetric space-time codes. Recent fast decodable code constructions include a fully diverse MIDO code conjectured to have the NVD property \cite{SR}, and MIDO codes with the NVD property based on crossed product algebras \cite{OVH,LO}.
Constructions for arbitrary number of transmit antennas are also available in \cite{NR2}.

%%%%%%%%%%%%%%%%%%%%%%%%%%%%%%%%%%%%%%%%%%%%%%%%%%%%%%%%%%%%%%%%%%%%%%%%%%%%%%%%%%%%%%%%%%%
\subsection{System Model and Fast Decodability}

We consider transmission over a coherent Rayleigh fading channel with $2n$ transmit antennas, $n$ receive antennas and perfect channel state information at the receiver (CSIR):
\begin{equation}\label{eq:channel}
Y_{n\times 2n}=H_{n\times 2n}X_{2n\times 2n}+V_{n\times 2n},
\end{equation}
where $H$ is the channel matrix and $V$ is the noise at the receiver, both with
i.i.d zero mean complex Gaussian components.
The $2n\times 2n$ space-time code $X$ can transmit up to $2n^2$ complex (say QAM) information
symbols, or equivalently $4n^2$ real (say PAM) information symbols. We will focus on the full-rate case where indeed $4n^2$ real symbols are sent. When $n=2$, $X$ is sometimes referred to as MIDO space-time code, where MIDO stands for multiple input double output.

Maximum-likelihood (ML) decoding consists of finding the codeword $X$ that
achieves the minimum of the squared Frobenius norm
\begin{equation}
\label{frob-min}
d(X)=||Y_{n\times 2n}-H_{n\times 2n}X||_F^2.
\end{equation}
By describing the space-time code $X$ in terms of basis matrices
$B_i$, $i=1,\ldots,4n^2,$ and a PAM vector $\gv=(g_1,\dots,g_{4n^2})^T$ as
\begin{equation}\label{eq:Xencod}
X=\sum_{i=1}^{4n^2} g_i B_i,
\end{equation}
we can rewrite (\ref{frob-min}) in Euclidean norm as
\begin{equation}\label{eq:dlatt}
d(X) = || \yv -B\gv ||_E^2,
\end{equation}
where $\yv\in\RR^{4n^2}$ is the channel output which has been vectorized, with real and imaginary parts separated, and
\begin{equation}\label{matrixB}
B=(\bv_1, \bv_2,\dots, \bv_{4n^2}) \in M_{4n^2\times 4n^2}(\RR),
\end{equation}
also obtained by vectorizing and separating the real and imaginary parts of $H_{n\times 2n}B_i$ to obtain
$\bv_i,~i=1,\ldots,4n^2$.
This search can then be performed using a real sphere decoder \cite{VB}.

We will focus on two aspects of space-time code design: diversity and fast decodability.

{\bf Full diversity.}
It is well known \cite{Tarokh} that the first important space-time code design criterion is {\emph{full diversity}}, that is, we require $\det(X-X') \neq 0$, for any $X\neq X'$ in the codebook. 
When $X\pm X'$ is again a codeword for all $X,X'$, the full diversity criterion simplifies to
\begin{equation}\label{eq:fulldiv}
\det(X)\neq 0 {\textrm{ for all}}~X\neq 0.
\end{equation}

{\bf Fast decodability.}
From (\ref{eq:dlatt}), a QR decomposition of $B$, $B=QR$, with $Q$ unitary, reduces to computing
\begin{equation}\label{eq:dR}
d(X)=|| \yv-QR\gv||_E^2=||Q^*\yv-R\gv||_E^2
\end{equation}
where $R$ is an upper right triangular matrix and $()^*$ denotes the Hermitian transpose. The number and position of
nonzero elements in the upper right part of $R$ will determine the complexity
of the sphere decoding process.
\begin{defi}
If $S$ is the real alphabet in use, and $\kappa$ is the number of independent real information symbols from $S$ within one code matrix, then the {\em ML decoding complexity}~\cite{BHV} is the minimum number of values of $d(X)$ in
(\ref{eq:dR}) that should be computed while performing  ML decoding.
\end{defi}
Note that $\kappa$ will sometimes be referred to as the rank or dimension of the code, which corresponds to the number of basis matrices, $\kappa=4n^2$ in our case.
The worst case is given when the matrix $R$ is a full upper right
triangular matrix, yielding the complexity of the exhaustive-search ML
decoder, that is here $O(|S|^{4n^2})$, with $\kappa=4n^2$ and $|S|$ is the number of PAM symbols in use.
If the structure of the code is such that the decoding complexity has an exponent of $|S|$ smaller than $4n^2$, we say that the code is {\em fast-decodable}.

We now recall three shapes of matrix $R$ which have been shown \cite{JR, NR} to result in reduced decoding complexity. 

Suppose a space-time code has matrix $R$ of the form:
$$R = {\bm {\Delta} {B_1} 0 {R_2}},$$
where $\Delta$ is a $d\times d$ diagonal matrix and $R_2$ is upper-triangular. Then decoding complexity of the code is reduced to $O(|S|^{\kappa-d+1})$. 
The matrix $R$, which depends on the channel matrix $H$, may be difficult to compute. Thankfully, the zero structure of $R$ is actually related \cite{JR} to the zero structure of the matrix $M$ defined by
\begin{equation}
M_{k,l}= ||B_kB_l^*+B_lB_k^*||_F,
\label{orthRelations}
\end{equation}
which captures information about orthogonality relations of the basis elements $B_i$, with $B^*_i$ denoting Hermitian transpose. It furthermore has the advantage that its zero structure is stable under premultiplication of $B_i $ by a channel matrix $H$ (in general, the same does not hold for $R$). It was shown in \cite[Lemma 2]{JR} that
\[
M = \bm {\Delta} {B_1} {B_2} {B_3} \Rightarrow R = \bm {\Delta} {B_1} {0} {R_1},
\]
where $\Delta$ is a $d$-dimensional diagonal matrix.
It is thus sufficient to compute $M$ to deduce fast-decodability. 

\begin{defi}
A space-time code of dimension $\kappa$ is called {\emph{g-group decodable}} if there exists a partition of $\{1, \ldots, \kappa\}$ into $g$ nonempty subsets $\Gamma_1, \ldots, \Gamma_g$, so that $M_{k,l}=0$ when $k, l$ are in disjoint subsets $\Gamma_i , \Gamma_j$. \end{defi}

In this case, as shown in \cite{JR}, the matrix $R$ has the form $R = diag(R_1, \ldots, R_g)$, where each $R_i$ is a square upper triangular matrix. Hence, the symbols $g_k$ and $g_l$ can be decoded independently when their corresponding basis matrices $B_k$ and $B_l$ belong to disjoint subsets of the induced partition of the basis, yielding a decoding complexity of $O(|S|^{\max_{1 \leq i \leq g} {\Gamma_i}})$.

Finally, a code might not be a $g$-group decodable code, but could become one assuming that a set of symbols are already decoded. This corresponds to another characterization of fast-decodability \cite{NR}:
\begin{defi}
A code is called \emph{conditionally $g$-group decodable} if there exists a partition of $\{1,\ldots,\kappa\}$ into $g+1$ disjoint subsets $\Gamma_1$, \ldots, $\Gamma_g$, $\Gamma^C$ such that
$$\|B_lB_m^*+B_mB_l^* \|_F =0 \quad \forall l \in \Gamma_i, \forall m \in \Gamma_j, i \neq j.$$
\end{defi}
In this case, the sphere decoding complexity order reduces to $|S|^{{\Gamma^C}+\max_{1 \leq i \leq g} {\Gamma_i}}$.

By the above discussion, in order to demonstrate fast-decodability, respectively (conditional) $g$-group decodability, it suffices to find an ordering on the basis elements $B_i$, which results in the desired zero structure of $M$.

%%%%%%%%%%%%%%%%%%%%%%%%%%%%%%%%%%%%%%%%%%%%%%%%%%%%%%%%%%%%%%%%%%%%%%%%%%%%%%%%%%%%%%%%%%%
\subsection{Contribution and Organization}

In this paper, we consider the problem of designing full-rate space-time codes for $2n$ transmit and $n$ receive asymmetric MIMO channels.
Our main contributions are:
\begin{enumerate}
\item
We propose a general code construction that maps two { {rank-$n$}} algebraic space-time codes coming from a cyclic algebra to a  { {rank-$2n$}} new code and give a criterion for full diversity. 
\item 
We obtain a novel way of  constructing division algebras, and describe their center and { {one}} maximal subfield.
\item
We provide a decoding complexity analysis that permits to derive conditions under which the iterated code inherits fast-decodable properties of the underlying code.
\item
We construct iterated MIDO versions of the well known Golden \cite{BRV} and Silver code 
\cite{HLRVV}, as well as MIMO codes for the much less studied case of $6$ transmit and $3$ receive antennas.
\end{enumerate}
This paper is organized as follows. The main construction is explained in Section \ref{construction}, together with its main properties - full diversity, algebra structure and fast decodability. We focus on the { { case of  quaternion algebras}} in Section \ref{sec:quat}, where we reformulate full diversity in terms of quadratic forms. MIDO codes constructions are provided in Section \ref{sec:MIDO}, where we construct iterated versions of well known quaternion based codes, such as the Silver and Golden code. Finally in Section \ref{sec:higher}, cyclic algebras of degree 3 are investigated. 
Simulation results illustrate the performance of the codes.

%***********************************************************************************%
%
% GENERAL CONSTRUCTION
%
%***********************************************************************************%
\section{An Iterated Code Construction}
\label{construction}

Let $F$ be a finite Galois extension of the field $\Q$ of rational numbers.
Let $K/F$ be a cyclic Galois extension of degree $n$ with Galois group generated by $\sigma$. 
Let $\mathcal D := (K/F, \sigma, \gamma) = K \oplus eK \oplus \cdots \oplus e^{n-1}K $ be the cyclic central simple algebra over $F$ defined by the relations
$$e^n = \gamma \in F, $$
$$ le = e \sigma(l) ~\forall l \in K.$$

{ {As a vector space, $\Dc$ is of dimension $n^2$ over its center $Z(\Dc) = F$. The algebra $\Dc$ is said to be {\emph{of  degree}} $n$ over its center $F$.
}}
We say that $\Dc$ is {\emph{division}} when every nonzero element of $\Dc$ is invertible.
 
{ {Left multiplication of elements of $\Dc$ induces the representation  $\lambda$ in $M_{n\times n}(K)$,}} which maps elements of $\Dc$ into the ring $M_{n \times n}(K)$ of $n\times n$ matrices with entries in $K$ \cite{SRS}, given by
{ {
\[\lambda: l \mapsto \begin{bmatrix}
l & 0  &0  &0 \\
0 & {\sigma(l)} &0  &0       \\
  & & \ddots  &   \\
0&0&0& \sigma^{n-1}(l)  \ \\
\end{bmatrix},   l \in K \]
\[\lambda: e \mapsto
\begin{bmatrix}
0 &   &  & \gamma \\
1 & 0 &  &       \\
  && \ddots  &   \\
  &   &  1&  0 \\
\end{bmatrix}.\]
}}

{ {
Any automorphism $\tau$ of $K$ extends naturally to an automorphism of $M_{n\times n}(K)$, also denoted by $\tau$
$$\tau: (A_{ij}) \mapsto (\tau(A_{ij}))$$ by applying $\tau$ on the matrix coefficients. 
Additionally assume the following: 
 \begin{equation}
 \tau(\gamma)=\gamma, ~ \tau\sigma = \sigma \tau
 \label{assumption1}
 \end{equation} 
 Then the automorphism $\tau$ of $M_{n\times n}(K)$ restricts to an automorphism of $\lambda(\Dc)$. 
Alternatively, the automorphism $\tau$ of $K$ extends to the automorphism $\tau$ of $\Dc$:
$$\tau: x_0+e x_1 + \cdots + e^{n-1} x_{n-1} \mapsto $$
$$\tau(x_0)+e \tau(x_1)+\ldots + e^{n-1} \tau(x_{n-1})$$
and hence $\tau$ extends to { {the  representation $\lambda(\Dc)$ via $\tau: \lambda(z)\mapsto \lambda(\tau(z))$. 
Putting the above together we have the following commutative diagram \[
\begin{diagram}
 \node{\Dc}\arrow{e,r}{\lambda}\arrow{s,l}{\tau}  \node{\lambda(\Dc)}\arrow{s,l}{\tau}\arrow{e,l}{} \node{M_{n\times n}(K)}\arrow{s,r}{\tau} \\
 \node{\Dc}\arrow{e,r}{\lambda} \node{\lambda(\Dc)}\arrow{e,l}{}   \node{M_{n\times n}(K).} 
\end{diagram}
\]
}}
}}

Note that for matrices $A, B \in \lambda(\Dc)$, $\tau(A)\tau(B) = \tau(AB)$. Hence for $A$ invertible, we have $\tau(A^{-1})= \tau(A)^{-1}$.

%%%%%%%%%%%%%%%%%%%%%%%%%%%%%%%%%%%%%%%%%%%%%%%%%%%%%%%%%%%%%%%%%%%%%%%%%%%%%%%%%%%%%%%%%%%%%%%%%%%%%%%%%%%%%%
\subsection{The General Code Construction}
{ {
Given an element $\theta$ of $\Dc$ and a $\Q$-automorphism $\tau$ of $K$, let 
$$\alpha_{\theta}: M_{n\times n}(K)\times M_{n\times n}(K) \rightarrow M_{2n \times 2n}(K)$$ be the map defined by
\begin{equation}\label{eq:alpha}
\alpha_{\theta} : (X,Y) \mapsto
{\begin{bmatrix}
{X} & {\theta \tau(Y)} \\ Y& {\tau(X)}
\end{bmatrix}}.
\end{equation}
If in addition the assumptions in (\ref{assumption1}) hold, then the restriction of $\alpha_{\theta}$ to $\lambda(\Dc)\times \lambda(\Dc)$ can be seen as an embedding of $\Dc \times \Dc$ into $M_{2n \times 2n}(K)$ 
by identifying elements $x,y \in \Dc$ with their representation $\lambda(x)=X, \lambda(y)=Y \in M_{n\times n}(K)$. 
}}
\begin{remark}
In fact, this construction can be made more general in two ways: (1) one could start with a noncyclic  central simple algebra  (this has been proposed in \cite{mtns}), and (2) the map $\alpha_\theta$ could be applied iteratively to obtain a family of codes for $2^i$ transmit antennas, $i > 0$. 
\cite{MOisit}. 
\end{remark}
{ {
\begin{remark}
Note that the  case when $\tau \neq \sigma$ and both are Galois automorphisms of $K$ of order $2$, the iterated construction coming from $\alpha_\theta(\Dc, \Dc)$ is in fact the same as the matrix representation of the crossed product $(K/F, \gamma, \theta, 1)$ \cite{BO} induced by left multiplication, where  the automorphisms $\sigma, \tau$ generate the biquadratic extension $K/F$ and $\Dc = (K/K^{\<\sigma\>}, \sigma, \gamma)$. 
\end{remark}
}}
We start with a simple but important criterion to decide when the image $\A$ of $\alpha_\theta$ has the structure of  { {a finite dimensional $\Q$-algebra.}}
\begin{lemma}\label{AisAlgebra}
{ {Let $\Dc = (K/F, \sigma, \gamma)$ and let $\tau \in Gal(K/F)$ be an automorphism such that assumptions in (\ref{assumption1}) hold, i.e., $\tau \sigma = \sigma \tau$ and $\tau(\gamma)=\gamma$. 
Additionally, suppose $\tau^2=1$ and fix $\theta \in F$ such that $\tau(\theta)=\theta$. Then the image  $\A = \alpha_{\theta}(\Dc, \Dc)$ forms a $\Q$-algebra. 
}}
\end{lemma}
\begin{IEEEproof}
{ {
Note that the assumptions $\tau \sigma = \sigma \tau$ and $\tau(\gamma) = \gamma$ allow us to extend the automorphism $\tau$  from $K$ to $\Dc$. 
}}
Next observe that the image $\A$ of $\alpha_{\theta}$ is both additively and multiplicatively closed. The additive closure is immediate: 
$$
\alpha_\theta(x,y)+\alpha_\theta(u,v)=\begin{bmatrix}
x & \theta \tau(y) \\ y & \tau(x)\\
\end{bmatrix}+
\begin{bmatrix}
u & \theta \tau(v) \\ v & \tau(u)\\
\end{bmatrix}$$
$$
=\alpha_\theta(x+u,y+v).
$$
For the multiplicative closure, we verify that 
$$
\alpha_\theta(x,y)\alpha_\theta(u,v) =
\begin{bmatrix}
xu+\theta \tau(y)v & 
{ {x \theta \tau(v)+ \theta \tau(y)\tau(u) }}\\
yu+\tau(x)v  & 
{ {y\theta  \tau(v)+\tau(x)\tau(u)}}\\
\end{bmatrix}$$
$$
=
\alpha_\theta(xu +\theta \tau(y)v, yu+\tau(x)v)
$$
using { {the assumptions that $\tau(\theta)=\theta$, $\theta \in Z(\Dc)=F$ and $\tau^2=1$.}}
{ {Hence $\A$ is a ring. Since $\A$ is additionally a vector space over $\alpha_\theta(\Q, 0)$, which  is central in $\A$, identifying $\Q$ with $\alpha_\theta(\Q, 0)$, gives the ring $\A$  the structure of a $\Q$-algebra of (finite)  dimension $[\A:\Q] = 2[\Dc: F][F:\Q] = 2n^2[F:\Q]$. }}
%It is of dimension $2n^2$ over $F$, since $\Dc$ is of dimension $n^2$ over $F$.
\end{IEEEproof}

We now give a condition on $\theta$ which guarantees that the image $\A$ of $\alpha_{\theta}$ satisfies the full diversity property (\ref{eq:fulldiv}), irrespectively of whether $\A$ is an algebra (more precisely, $\A$ might not be multiplicatively closed).
\begin{lemma}\label{normConditionN}
Let $K/F$ be a cyclic extension of number fields and $\Dc = (K/F, \sigma, \gamma)$ be a cyclic division algebra and
{fix $\theta \in \Dc$.  }
 Let $\tau$ be a Galois $\Q$-automorphism of $K$ satisfying assumptions in (\ref{assumption1}), that is, 
$ \tau(\gamma)=\gamma, ~ \tau\sigma = \sigma \tau$. { {Every matrix in the image $\mathcal A$ of $\alpha_\theta$ has determinant in $F$ and is invertible
 if and only if $\theta \not = z\tau(z) $ for all $z \in \Dc$. 
{ { If moreover $\tau^2=1$ and $\theta \in F$, then  for any $\alpha_\theta(x,y) \in \mathcal A$, the determinant $\det(\alpha_\theta(x,y))$ is contained in the subfield $F^{\<\tau \>}$ of $F$ fixed by $\tau$. }}
}}

\end{lemma}
\begin{IEEEproof}
Suppose that $\theta \not = z\tau(z) $ for all $z \in \Dc$. We will show that $\mathcal A$ is fully diverse, { {i.e., every nonzero element of $\Ac$ has nonzero determinant.}}

Consider a nonzero element $\alpha_\theta(x,y) = \begin{bmatrix}
x & {\theta \tau(y)} \\ y & {\tau(x)}
\end{bmatrix}$, where the entries $x,y$ are $n\times n$ matrices from the division algebra $\Dc$. It is enough to demonstrate that the determinant of $\alpha_\theta(x,y)$ is nonzero, since $\Ac$ is additively closed. If $x=0$ (resp. $y=0$), the matrix is clearly invertible. Hence for $x,y \in \lambda(\Dc)$, we assume that  $x$ and $y$ are both nonzero, and hence, as elements of a division algebra, both invertible.
The determinant of a block matrix
$\begin{bmatrix}
A & B \\ C & D
\end{bmatrix} $, when $A$ is invertible, is given by $\det(A)\det(D-CA^{-1}B)$. Then 
%$M = \begin{bmatrix} x & {\theta \tau(y)} \\ {y} & {\tau(x)} \end{bmatrix} \in \A$, 

we have 
\begin{equation}
\det \begin{bmatrix} x & {\theta \tau(y)} \\ {y} & {\tau(x)} \end{bmatrix} = \det(x)\det(\tau(x)-yx^{-1}\theta \tau(y)).
\label{eq:det1}
\end{equation}
Note that $\tau(\gamma)=\gamma$, together with the fact that $\tau$ and $\sigma$ commute, imply that $\tau(x)\in\lambda(\Dc)$.
The assumption $\theta\in F$ is needed so that $\theta\tau(y)\in\lambda(\Dc)$. It then follows that  $\tau(x)-yx^{-1}\theta \tau(y)$ is an element of $\lambda(\Dc)$,
and hence invertible when nonzero: we have
$ \det(\tau(x)-yx^{-1}\theta \tau(y)) \neq 0  \iff \tau(x)-yx^{-1}\theta \tau(y) \neq 0. $
It thus  suffices to demonstrate
$\tau(x)-yx^{-1}\theta \tau(y) \neq 0. $
Since $y$ is invertible (it is nonzero), and
noting that for $x\in \lambda(\Dc)$, $\tau(x^{-1}) = \tau(x)^{-1}$, the latter inequality is equivalent to
$ xy^{-1}\tau(xy^{-1}) \neq \theta. $
Letting $z = xy^{-1}$, by our assumption on $\theta$, we have $\theta \neq z\tau(z)$. Hence $\det(\alpha_\theta(x,y)) \neq 0$ and is well known to belong to the center $F$ of $\Dc$, since both $x$ and $\tau(x)-yx^{-1}\theta \tau(y)$ are elements of $\Dc$.

{ {
Now assume additionally that  $\tau^2=1$ and $\theta \in F$.
We show that $\det(\alpha_\theta(x,y)) \in F^{\<\tau\>}$. Since by above remarks we already know that the determinant is in $F$, it suffices to show that it is fixed by $\tau$. Now when $D$ is invertible, we can alternatively express the determinant of the block matrix 
$\begin{bmatrix}
A & B \\ C & D
\end{bmatrix} $
as $\det(D)\det(A-BD^{-1}C)$. Using the assumption that $\tau(A)$ is invertible with thus get
\begin{equation}
\det \begin{bmatrix} x & {\theta \tau(y)} \\ {y} & {\tau(x)} \end{bmatrix} 
= \det(\tau(x))\det(x- \theta \tau(y){\tau(x)}^{-1}y) 
%&=& \det(x)\det(\tau(x)-yx^{-1}\theta \tau(y))
\label{eq:det2}
\end{equation}
Equating formulas (\ref{eq:det1}) and (\ref{eq:det2}) makes it clear that $\det(\alpha_\theta(x,y))$ is fixed by $\tau$. Hence we have $\det(\alpha_\theta(x,y)) \in F^{\<\tau\>}$. 
}}

Conversely, suppose $\theta = z \tau(z)$ for some $z \in \lambda(\Dc)$. We verify that $\det(\alpha_\theta(z,I_n))=0$, where $I_n$ is the $n$-dimensional identity matrix.  Indeed
\begin{eqnarray*}
\det(\alpha_\theta(z,I_n))&=& \det(z)\det(\tau(z)-I_nz^{-1} \theta) \\
&=& \det(z)\det(\tau(z)-I_nz^{-1} z\tau(z))\\
& =& \det(z)\det(0)= 0.
\end{eqnarray*}
Hence $\mathcal A$ is not fully diverse.
\end{IEEEproof}

\begin{corollary}\label{cor:adiv}
{ {
Let $\Dc=(K/F,\sigma,\gamma)$ be a cyclic division algebra. Consider a Galois $\Q$-automorphism $\tau$ of $K$,  such that assumptions (\ref{assumption1}) hold, i.e.,  
$ \tau(\gamma)=\gamma, ~ \tau\sigma = \sigma \tau$. Assume additionally that $\tau^2 = 1$ and fix $\theta \in F$ such that $\tau(\theta)=\theta$. 
Then the image $\Ac$ of $\alpha_{\theta}$ is an algebra, which is division if and only if
$\theta \not = z\tau(z) $ for all $z \in \Dc$.
}}
\end{corollary}

\begin{IEEEproof}
{ 
{
The assumptions ({\ref{assumption1}}) on $\tau$ allow to extend $\tau$ to an automorphism of $\Dc$, while the additional assumptions that $\tau^2 = 1, \theta \in F$ and $\tau(\theta)=\theta$ give the image $\Ac$ the structure of a finite dimensional $\Q$-algebra by Lemma \ref{AisAlgebra}. A finite dimensional algebra is division (i.e. every nonzero element has an inverse) if and only if it contains no zero divisors. If 
$\theta \not = z\tau(z) $ for all $z \in \Dc$, then by previous Lemma every element of $\A$ is an invertible matrix, and hence cannot be a zero divisor. Hence $\A$ is division (note that we avoided having to explicitly demonstrate that the inverse of every element is indeed in $\Ac$). On the other hand if 
$\theta  = z\tau(z)$ for some $z \in \Dc$, then by previous Lemma $\A$ contains a non-invertible matrix, and therefore cannot be division.  
}}
\end{IEEEproof}

%%%%%%%%%%%%%%%%%%%%%%%%%%%%%%%%%%%%%%%%%%%%%%%%%%%%%%%%%%%%%%%%%%%%%%%%%%%%%%%%%%%%%%%%%%%%%%%%%%%%%%%%%%%%%%
\subsection{More Algebraic Properties and Fast Decodability}

We saw in Lemma \ref{AisAlgebra} that assuming that $\tau^2=1$ and $\theta \in F^{\<\tau\>}$, the image $\Ac$ is an algebra. It is natural to wonder what is the center $Z(\A)$ of $\A$.

\begin{lemma} \label{ImageAlgebra}
{ {
Let $\tau$ be a Galois automorphism of $K$ of order $2$. 
If $\tau \in \<\sigma \>$, then the ring $\alpha_\theta(F, e^{n/2}F)$ is a maximal subring which commutes with all elements of $\A$.
Otherwise, $\alpha_\theta(F^\tau,0)$ is such a ring.}}
\end{lemma}

\begin{IEEEproof}
The Galois automorphism $\tau$ may or may not be an element of the Galois group $\langle \sigma \rangle$.  We treat the two cases separately. \\
\noindent{\bf{Case 1.}}
Let us first suppose the former case when $\tau \in \langle \sigma \rangle$. Since $\tau$ is of order $2$, it follows that $\tau=\sigma^{n/2}$.

Suppose that $\alpha_\theta(u,v)$ commutes with all the elements of $\A$, that is,  $\alpha_\theta(u,v)\alpha_\theta(x,y)=\alpha_\theta(x,y)\alpha_\theta(u,v)$
for any $x,y\in \Dc$. Then, in particular, it must be  true for $y=0$. This implies that $ux=xu$ and $\tau(x)v=vx$ must hold for all $x\in \Dc$. The first condition implies that $u \in Z(\Dc) = F$. The second implies that $\sigma^{n/2}(x)v=vx$. 

{ {
We show that this condition forces $v$ to be an element in $e^{n/2}F$. 
Express $v = v_0  + ev_1+ \cdots + e^{n-1}v^{n-1}$, where $v_1, \ldots, v_{n-1} \in K$. Now for all $x \in K$, we have 
$$\sigma^{n/2}(x)v= \sigma^{n/2}(x)(v_0  + ev_1+ \cdots + e^{n-1}v_{n-1}) = $$
$$v_0 \sigma^{n/2}(x) + ev_1 \sigma^{n/2+1}(x)+ \cdots + e^{n-1}v_{n-1}\sigma^{n/2+{n-1}}(x).$$ 
Equating with 
%$$v_0 \sigma^{n/2}(x) + ev_1 \sigma^{n/2+1}(x)+ \cdots + e^{n-1}v_{n-1}\sigma^{n/2+{n-1}}(x) = $$
$vx = v_0x  + ev_1x+ \cdots + e^{n-1}v_{n-1}x $
in each component gives 
$e^iv_i\sigma^{n/2+i}(x) = e^iv_i x$  for all $x\in K$. This forces for all  $v_i \neq 0$ the condition $\sigma^{n/2+i}(x) =  x$ for all    $ x \in K$. The latter is true if and only if $i=n/2$, forcing  $v_i$ for all $i \neq n/2$ to be zero. 
Hence we obtain $v \in e^{n/2}K$.
Now letting $x=e$ and writing $v=e^{n/2}v_{n/2}$, the constraint $\sigma^{n/2}(e)v=ve$, gives  
$e\cdot e^{n/2}v_{n/2} = e^{n/2}v_{n/2}e = e^{n/2}e\sigma(v_{n/2})$, implying that $v_{n/2}=\sigma(v_{n/2})$, i.e., $v \in e^{n/2}F$, as desired. 
}}
%Indeed, we verify that $e^{n/2}$ satisfies $\sigma^{n/2}(x)e^{n/2}=e^{n/2}x$ for all $x \in \Dc$. Hence $\alpha_\theta(u,v) \in \alpha_\theta(F,e^{n/2}F)$. 
On the other hand, we check that $\alpha_\theta(F,e^{n/2}F)$ in fact commutes with all elements of $\A$. This completes the first case.\\ 
\noindent{\bf{Case 2.}}
Now suppose that $\tau$ does not belong to $\langle \sigma \rangle$. 
If $\alpha_\theta(u,v)$ commutes with all the elements of $\A$, the conclusion that $u\in F$ holds similarly to  Case 1. 

Now we additionally need to impose the constraint $\tau(x)v=vx$ for all $x \in \Dc$. 
By writing $v = \sum_{i=0}^{n-1}e^iv_i$, this yields
{ { for any $x \in K$ the condition 
$  \sigma^i(\tau(x))v_i = v_ix$ for all $i$.  
Since $\sigma^{i}\tau$ is not identity on $K$ for any $i$, we conclude that $v_i = 0$ for all $i$,  showing that $v = 0$. 
}}
Next { {we check for which values of $u$ the element }}$\alpha_\theta(u,0)$ commutes with all elements $\alpha_\theta(0,y)$ where $y \in \Dc$. This { {produces the constraint}} $\tau(u)y = yu$ for all $y \in \Dc$, that is, $u \in F^\tau$. On the other hand,  $\alpha_\theta(F^\tau,0)$ indeed commutes with all the elements of $\A$, completing the proof. 
\end{IEEEproof}

\begin{remark}
In Corollary \ref{cor:adiv} we derived a criterion which forces $\A$ to be division. When this criterion is satisfied, then in case $\tau \in \<\sigma\>$, we  have $F':=\alpha_\theta(F,e^{n/2}F)$ is necessarily a field and it corresponds to the center of $\A$, while $K':=\alpha_\theta(K,e^{n/2}F)$ is a maximal subfield of $\A$. In fact, we can see that $\A \cong (K'/F', \sigma, \gamma)$ and is generated as a vector space over $K'$ by powers of the element $\alpha_\theta(e,0)$.
In case when $\tau \not \in \<\sigma\>$, the center of $\A$ is $\alpha_\theta(F^\tau, 0)$ and $\alpha_\theta(K,0)$ { {is a maximal commutative subfield of $\A$}}. The latter follows from the observation that $\alpha_\theta(K,0)$ is clearly commutative and  its maximality follows from the fact that it is of correct degree $2n$ over the center.

\label{centerA}\end{remark}
\begin{figure}
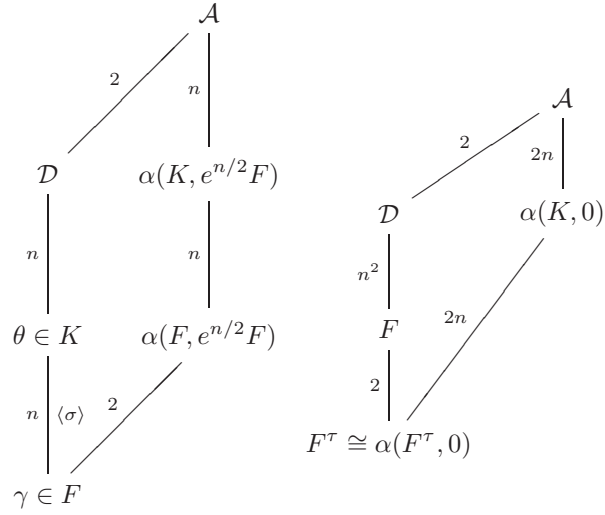

{ {
\[
%\begin{diagram}
 %\node{\Dc\simeq\oplus_{i=0}^{n-1} e^iK} \\
 %\node{K \ni \theta} \arrow{n,l,-}{n}\\
 %\node{F\ni \gamma} \arrow{n,r,-}{\langle \sigma \rangle} \arrow{n,l,-}{n} \\
%\end{diagram}
\begin{diagram}
 \node{} \node{\A}\arrow{s,l,-}{n} \\
 \node{\Dc} \arrow{ne,l,-}{2}\node{\alpha(K,e^{n/2}F)}\arrow{s,l,-}{n}\\
 \node{\theta \in K}\arrow{n,l,-}{n} \node{\alpha(F,e^{n/2}F)}\\
 \node{\gamma \in F} \arrow[1]{n,l,-}{n} \arrow{n,r,-}{\langle \sigma \rangle} \arrow{ne,l,-}{2} \\
\end{diagram}
\begin{diagram}
 \node{} \node{\A}\arrow{s,l,-}{2n} \\
 \node{\Dc} \arrow{ne,l,-}{2}\node{\alpha(K,0)}\arrow{ssw,l,-}{2n} \\
 \node{F} \arrow{n,l,-}{n^2}\\
\node{F^{\tau} \cong \alpha(F^{\tau}, 0)}\arrow{n,l,-}{2} \\
\node{}
\end{diagram}
\vspace{-1cm}
\]
}}
\caption{\label{fig:tower1}Tower of extensions showing the iterated construction: on the left, $\Dc$ is of dimension $n^2$ over its center $F$. In the middle, when $\tau\in\<\sigma\>$, $\A$ is of dimension 2 
over $\Dc$, that is of dimension $2n^2$ over $F$. It is further of dimension $n^2$ over the algebra $\alpha(F,e^{n/2}F)$, $e^n=\gamma$, which is itself of dimension 2 over $F$. On the right, when 
$\tau \not\in\<\sigma\>$, the algebra $\A$ is of dimension $4n^2$ over $F^\tau$. 
}
\end{figure}

The embedding $\alpha_{\theta}: \Dc\times \Dc \rightarrow M_{2n}(K)$  always has an additively closed image $\A$, which forms a vector space over $F$ of dimension $2n^2$. In fact, we can generate $\A$ { {over $F$}} by elements of the form
\begin{equation}\label{eq:BFD}
\{ \alpha_{\theta}(u,0), \alpha_{\theta}(0,u) : u = u_1, \ldots u_{n^2} \in \B_F(\Dc) \},
\end{equation}
where $\B_F(\Dc)$ is an $F$-basis of the cyclic algebra $\Dc = (K/F,\sigma,\gamma)$:
$$ \Dc = K\oplus eK \oplus \ldots \oplus e^{n-1}K,~e^n=\gamma \in F$$
where
\[
K=\nu_1 F\oplus \nu_2 F \oplus \ldots \oplus \nu_n F
\]
by fixing an $F$-basis $\{\nu_1,\ldots,\nu_n \}$ of $K$.
We can thus choose
\[
\B_F(\Dc)=\{ \nu_1,\ldots,\nu_n,e\nu_1,\ldots,e\nu_n,\ldots,e^{n-1}\nu_1,\ldots,e^{n-1}\nu_n\}
\]

{ {
If $F$ is a quadratic field extension, we can obtain a $\QQ$-basis $\B_{\QQ}(\Dc) = \{D_j,~j=1,\ldots, 2n^2\}$ of $\Dc$ using a $\QQ$-basis $\{\beta_1, \beta_2\}$ of $F$: 
$$ \{D_1, \ldots, D_{n^2}, D_{n^2+1},\ldots,  D_{2n^2}\} 
= \beta_1 \B_{F}(\Dc) \cup \beta_2 \B_{F}(\Dc) $$
\[\{ \nu_1\beta_1,\ldots, e^{n-1}\nu_n\beta_1,  \nu_1\beta_2,\ldots, e^{n-1}\nu_n\beta_2 \} \label{qBasis}\]
%$$\{ \nu_1\beta_1,\ldots,\nu_n\beta_1,e\nu_1\beta_1,\ldots,e\nu_n\beta_1,\ldots,e^{n-1}\nu_1\beta_1,\ldots,e^{n-1}\nu_n\beta_1, \nu_1\beta_2,\ldots,\nu_n\beta_2,e\nu_1\beta_2,\ldots,e\nu_n\beta_2,\ldots,e^{n-1}\nu_1\beta_2,\ldots,e^{n-1}\nu_n\beta_2\}$$
When $F$ is a field of higher degree, we let $\beta_1, \beta_2$ be any two $\QQ$-independent elements of $F$, to define a $\QQ$-independent set 
$$\{D_1, \ldots, D_{n^2}, D_{n^2+1},\ldots,  D_{2n^2}\} = $$
$$\{ \nu_1\beta_1,\ldots, e^{n-1}\nu_n\beta_1,  \nu_1\beta_2,\ldots, e^{n-1}\nu_n\beta_2 \}.$$
}}
This finally yields 
{ {
a generating set
}}
\[
\B(\A) = \{ B_j,~j=1,\ldots,4n^2\}= 
\]
\[\{ \alpha_{\theta}(u,0), \alpha_{\theta}(0,u) : u = D_1, \ldots D_{2n^2}\}
\]
and encoding as described in (\ref{eq:Xencod}) consists of mapping a vector $(g_1,\ldots,g_{4n^2})$ of PAM information symbols (that is from $\mathbb{Z}$) into a space-time codeword in $\A$ using the above basis matrices:
$$
(g_1, \ldots, g_{4n^2})\mapsto X = \sum_{i=1}^{2n^2}(g_i\alpha_\theta(D_i,0)+g_{2n^2+i}\alpha_\theta(0,D_i)).
$$

We will use the following scaling technique.
Let $u,v, x, y \in \Dc$. { {When $\theta$ is either real or totally imaginary, we express $\theta=\zeta\theta'$ with $\theta'>0$, for some $\zeta\in\{\pm 1,\pm i\}$. Then the map}}
\[
\alpha_{\zeta\theta'}:(u,v)\mapsto
\begin{bmatrix}u & \zeta\theta'\tau(v) \\ v & \tau(u) \end{bmatrix}
\]
and the map $\tilde{\alpha}_{\zeta\sqrt{\theta'}}$ defined by
\begin{equation}\label{eq:alphatilde}
\tilde{\alpha}_{\zeta\sqrt{\theta'}}:(u,v)\mapsto
\begin{bmatrix}u & \zeta\sqrt{\theta'}\tau(v) \\ \sqrt{\theta'}v & \tau(u) \end{bmatrix}
\end{equation}
satisfy the property that $\det(\alpha_{\zeta\theta'}(u,v))=\det(\tilde{\alpha}_{\zeta\sqrt{\theta'}}(u,v))$, for all $u,v$. 
Note alternatively that 
\[
\tilde{\alpha}_{\zeta\sqrt{\theta'}}(u,v)=
\begin{bmatrix}1 & 0 \\ 0 & \sqrt{\theta'} \end{bmatrix}
\alpha_{\theta}(u,v)
\begin{bmatrix}1 & 0 \\ 0 & 1/\sqrt{\theta'} \end{bmatrix}
\]
and in particular, the image of $\tilde{\alpha}_{\zeta\sqrt{\theta'}}$ retains the full diversity (resp. NVD) property of the image of ${\alpha}_{\zeta{\theta'}}$.

Furthermore, assuming that complex conjugation commutes with $\tau$ on elements of $\Dc$,
we get 
$$
\tilde{\alpha}_{\sqrt{\theta'}}(u,v)^*=\begin{bmatrix}u & \zeta\sqrt{\theta'}\tau(v) \\ \sqrt{\theta'}v & \tau(u) \end{bmatrix}^*
= \begin{bmatrix} u^* & \sqrt{\theta'}v^* \\ \zeta^*\sqrt{\theta'}\tau(v^*) & \tau(u^*)  \end{bmatrix},
$$
so that letting $\tilde{\alpha}$ denote $\tilde{\alpha}_{\zeta\sqrt{\theta'}}$ we have
\begin{eqnarray*}
\tilde{\alpha}(x,y)\tilde{\alpha}(u,v)^*
=
\begin{bmatrix}x & \zeta\sqrt{\theta'} \tau(y) \\ \sqrt{\theta'}y & \tau(x) \end{bmatrix}
\begin{bmatrix} u^* & \sqrt{\theta'}v^* \\ \zeta^*\sqrt{\theta'}\tau(v^*) & \tau(u^*)  \end{bmatrix}\\
=
\begin{bmatrix}
xu^*+\theta'\tau(yv^*) & \sqrt{\theta'}(xv^*+\zeta\tau(yu^*)) \\
\sqrt{\theta'}(yu^*+\zeta^*\tau(xv^*)) & \theta' yv^*+\tau(xu^*)
\end{bmatrix}.
\end{eqnarray*}
Since $\zeta$ is a fourth root of unity, we notice that $xv^*+\zeta\tau(yu^*)=\zeta(\zeta^*xv^*+\tau(yu^*))$ and
\[
\tilde{\alpha}(x,y)\tilde{\alpha}(u,v)^*=\tilde{\alpha}(xu^*+\theta'\tau(yv^*),yu^*+\zeta^*\tau(xv^*)).
\]
This allows us to compute
{\small{
\begin{eqnarray}
\tilde{\alpha}(x,0)\tilde{\alpha}(u,0)^* =  \tilde{\alpha}(xu^*,0) = \begin{bmatrix} xu^* & 0 \\ 0 & \tau(xu^*) \end{bmatrix}\label{eq:mult1}\\
\tilde{\alpha}(x,0)\tilde{\alpha}(0,v)^* =  \tilde{\alpha}(0, \zeta^*\tau(xv^*))=\sqrt{\theta'}\begin{bmatrix} 0 & xv^* \\ \zeta^*\tau(xv^*) & 0\end{bmatrix} \label{eq:mult2}\\
\tilde{\alpha}(0,y)\tilde{\alpha}(u,0)^* =  \tilde{\alpha}(0, yu^*) = \sqrt{\theta'}\begin{bmatrix} 0 & \zeta\tau(yu^*) \\ yu^* & 0\end{bmatrix}
\label{eq:mult3}\\
\tilde{\alpha}(0,y)\tilde{\alpha}(0,v)^* = \tilde{\alpha}(\theta'\tau(yv^*),0) = \theta'\begin{bmatrix} \tau(yv^*) & 0 \\ 0 & yv^* \end{bmatrix}
\label{eq:mult4}
\end{eqnarray}
}}

from which we see that fast decodable properties of the code coming from $\Dc$ will be inherited by the image $\tilde{\alpha}_{\zeta\sqrt{\theta'}}(\Dc, \Dc)$. More precisely:

\begin{lemma}\label{lem:fd}
Let $\Dc$ be a division algebra whose corresponding code has basis $\{D_1,\ldots,D_{2n^2}\}$.
If
\[
D_jD_k^*+D_kD_j^*=0
\]
for some $j,k$, then the basis $\{B_1,\ldots,B_{4n^2}\}$ of the code coming from $\Ac=\alpha_{\theta}(\Dc,\Dc)$ given by 
\begin{equation}\label{eq:goodbasis}
\{ B_i=\tilde{\alpha}(D_i,0),~B_{2n^2+i}=\tilde{\alpha}(0,D_i),~i=1,\ldots,2n^2\}
\end{equation}
satisfies
\[
B_jB_k^*+B_kB_j^*=B_{2n^2+j}B_{2n^2+k}^*+B_{2n^2+k}B_{2n^2+j}^*=0.
\]
\end{lemma}
\begin{IEEEproof}
Indeed, let $\{D_1,\ldots,D_{2n^2}\}$ be a basis for the code $\Dc$ and (\ref{eq:goodbasis}) be one for the code from $\Ac$.
We need to compute $B_jB_k^*+B_kB_j^*$ for two groups of indices:  $j,k\in\{1,\ldots,2n^2\}$ and
$j,k\in\{2n^2+1,\ldots,4n^2\}$. In the first case, we use (\ref{eq:mult1}) to
see that
\[B_jB_k^*+B_kB_j^* = \]
\[
\begin{bmatrix} D_jD_k^* & 0 \\ 0 & \tau(D_kD_j^*) \end{bmatrix}+
\begin{bmatrix} D_kD_j^* & 0 \\ 0 & \tau(D_jD_k^*) \end{bmatrix} =\]
\[
\begin{bmatrix}D_jD_k^*+ D_kD_j^* & 0 \\ 0 & \tau(D_jD_k^*+D_jD_k^*) \end{bmatrix}
\]
from which it is clear that the orthogonal relations among $D_j$ determine those among $B_j$. 
The other equality is handled similarly, using (\ref{eq:mult4}).
\end{IEEEproof}
{ 
{Further improvements on these reductions in complexity may be possible for particular choices of $\zeta, \tau$ and are left for exploration in future work. 
}}
%***********************************************************************************%
%
% QUATERNION CONSTRUCTION
%
%***********************************************************************************%
\section{The case of Quaternion Algebras}
\label{sec:quat}

In this section we focus on the case when the underlying division algebra is of degree $2$, and can equivalently be viewed as a generalized quaternion algebra. Additionally, we will only consider the case when $\tau = \sigma$.

\begin{figure}
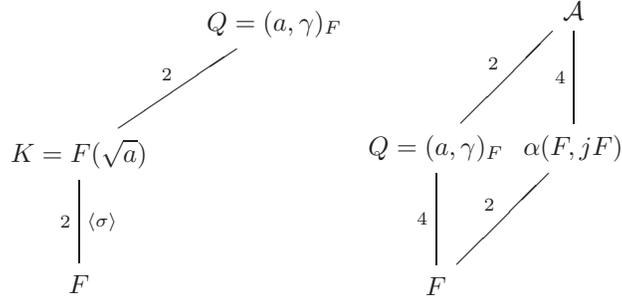

\[
\begin{diagram}
 \node{} \node{Q= \qa a \gamma F} \\
 \node{K=F(\sqrt a)} \arrow{ne,l,-}{2}\\
 \node{F} \arrow{n,r,-}{\langle \sigma \rangle} \arrow{n,l,-}{2} \\
\end{diagram}
\begin{diagram}
 \node{} \node{\A}\arrow{s,l,-}{4} \\
 \node{Q=\qa a \gamma F} \arrow{ne,l,-}{2}\node{\alpha(F,jF)}\\
 \node{F} \arrow{n,l,-}{4}\arrow{ne,l,-}{2} \\
\end{diagram}
\]
\caption{\label{fig:towerQ}Tower of extensions showing the iterated construction. On the left, $Q$ is of dimension 4 over its center $F$. On the right, $\A = \alpha_\theta(Q,Q)$ for $\tau = \sigma$ and $\theta \in F$, is of dimension 2 over $Q$,
that is of dimension 8 over $F$. It is further of dimension 4 over the subring $\alpha(F,jF)$, $j^2=\gamma$, which is itself of dimension 2 over $F$.
}
\end{figure}

{ {
Let $a,b\in F^\times$. We recall from \cite{UM} that a \emph{(generalized) quaternion algebra} over $F$, denoted by $\qa a b F$, is a $4$-dimensional $F$-vector space with $F$-basis $\{1, \vc i, \vc j, \vc k\}$, whose ring structure is determined by the rules
\begin{equation}\label{def.rel}
\vc i^2=a,\ \vc j^2=b\quad\text{and}\quad \vc{ij}=-\vc{ji}=\vc k.
\end{equation}
Let $F$ be a number field and let $Q=\qa a \gamma F$  be a generalized quaternion division algebra. 
We can alternatively  view $Q$ as a cyclic division algebra $(K/F, \sigma, \gamma)$, where a maximal subfield of $Q$ is a quadratic extension $K = F(\sqrt{a})$ of $F$ with Galois group generated by the automorphism $\sigma: \sqrt{a} \mapsto -\sqrt{a}$. }}
The diagram of Fig.~\ref{fig:towerQ} illustrates this tower of extensions.

The fact that $Q$ is a division algebra implies that $\gamma$ is not a norm of $K$ \cite{UM}.  We can write
$$
Q=K\oplus jK ,~j^2=\gamma \in F,
$$
and view elements of $Q$ as $2\times 2$ matrices over $K$ via { {matrix representation:}}
$$
c+ jd \mapsto
\begin{bmatrix}
c & \gamma \sigma(d) \\ d & \sigma(c)
\end{bmatrix}.
$$
The map $\alpha_{\theta}$ as defined in (\ref{eq:alpha}) with $\tau=\sigma$ is given here by
$$
\alpha_{\theta} : (x,y) \mapsto
\begin{bmatrix}
x & \theta \sigma(y) \\ y & \sigma(x)
\end{bmatrix}.
$$
By identifying $x$ and $y$ with their { {matrix representation}}, $\alpha_{\theta}$ can be seen as an embedding of $Q\times Q$ into $M_4(K)$. From Lemma \ref{AisAlgebra}, we have that $\Ac=\alpha_{\theta}(Q,Q)$ is an $F$-algebra whenever $\theta\in F$. 

\begin{example}\label{HQ}
Consider $Q = \qa {-1} {-1} \Q$ and write
$$
Q=\Q(i)\oplus j\Q(i) ,~j^2=-1.
$$

Note that $-1$ is not a norm of $\Q(i)$ over $\Q$, i.e., $-1 \neq (f+gi)(f-gi)$, for $f,g \in \Q$. 
Therefore $Q$ is a division algebra; its maximal subfield $\Q(i)$ is an extension of degree $2$ over its center $\Q$ with Galois group $Gal(\Q(i)/\Q)$ cyclic 
of order $2$. In this case, the generator $\sigma$ of the Galois group is the complex conjugation: $\sigma: x \mapsto \cc x$, $x\in \QQ(i)$.

We view the elements of $Q$ as $2\times 2$ matrices over $\Q(i)$ via { {matrix representation}}:
\begin{equation}\label{eq:alamcodeword}
c+ jd \mapsto
\begin{bmatrix}
c & - \cc d \\ d & \cc c
\end{bmatrix}.
\end{equation}
We illustrate the iterated construction arising from the quaternion algebra $Q = \qa {-1}{-1}{\Q}$ in Figure \ref{fig:quatalam} where $\theta\in\QQ$ is chosen so that $\Ac=\alpha_\theta(\Q, \Q)$ is division, by Corollary \ref{cor:adiv}. The center and  maximal subfield of $\A$ correspond to $\alpha_\theta(\Q, j\Q)$, $\alpha_\theta(\Q(i), j\Q(i))$, respectively, as follows from Lemma \ref{ImageAlgebra} and Remark 
\ref{centerA}. 

\begin{figure}
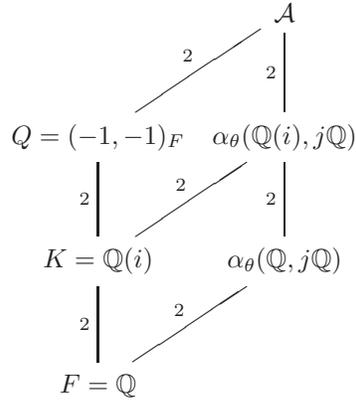

\[\begin{diagram}
 \node{} \node{\A}\arrow{s,l,-}{2} \\
 \node{Q=\qa {-1}{-1} F} \arrow{ne,l,-}{2}\node{\alpha_\theta(\Q(i),j\Q)}\\
 \node{K=\Q(i)}\arrow{n,l,-}{2}\arrow{ne,l,-}{2}\node{\alpha_\theta(\Q,j\Q)}\arrow{n,l,-}{2}\\
 \node{F=\Q} \arrow{n,l,-}{2}\arrow{ne,l,-}{2} \\   
\end{diagram}
\]
\caption{
The iterated construction from Example \ref{HQ} when $\Ac$ is division.
\label{fig:quatalam}
}
\end{figure}

\end{example}

%{ {The following example did not serving any purpose, deleted. }}

%%%%%%%%%%%%%%%%%%%%%%%%%%%%%%%%%%%%%%%%%%%%%%%%%%%%%%%%%%%%%%%%%%%%%%%%%%%%%%%%%%%%%%%%%%%%%%%%%%%%%%%%%%%%%%%%%%%%%
\subsection{First Criteria for Full Diversity}

The following lemma gives necessary and sufficient conditions on $\theta, \gamma$ in order to assure that the image $\A$ of $\alpha_\theta$  is a division algebra, following Corollary \ref{cor:adiv}. 
Let $K^{\times}$ denote the multiplicative group of nonzero elements in $K$.

\begin{lemma}\label{Condition}
Let $K = F(\sqrt{a}), \gamma, \theta \in F$. We have the following equivalence:

$$\theta \neq z\sigma(z) \text{ for any } z \in Q=\qa a \gamma F$$  $$\iff $$
\begin{enumerate}
\item
$ \theta \not = \det\bm{u} {\gamma \sigma(v)} { v }{ \sigma(u)},$ for any $u,v \in K$ such that $ v\in \sqrt{a}F,$ and
\item
$\theta \not = \gamma~(\mod K^{\times 2})$, 
{ {where $(\mod K^{\times 2})$ denotes taking the quotient by the subgroup $K^{\times 2}$ of squares in $K^\times$.
}}
\end{enumerate}
\end{lemma}
\begin{IEEEproof}
Assume conditions 1) and 2). For the sake of contradiction, suppose there exists an element $z\in Q$ such that $\theta = z\sigma(z)$, and write it as $z = \begin{bmatrix} u & \gamma\sigma(v)\\ v &\sigma(u) \end{bmatrix}$, where $u, v \in K$.

This gives
$$
\bm {u\sigma(u)+\gamma\sigma(v)^2}
{u \gamma v + \gamma \sigma(v)u}
{v\sigma(u)+\sigma(u)\sigma(v)}
{v\gamma v +\sigma(u)u}  = \bm \theta 0 0 \theta.
$$

Suppose $ u \not = 0$.
Comparing entries $(1,2)$ of the above matrices (keeping in mind that $u,v \in K$, and hence commute) gives $v+\sigma(v)=0,$
which implies that $v = v_1\sqrt{a}$, for some $v_1 \in F$, since $K$ is a quadratic extension of $F$ .
Comparing entries $(1,1)$ gives
$$u\sigma(u) + \gamma \sigma(v^2) = \theta. $$
Equivalently, recalling the previous condition that $v = v_1\sqrt{a}$, we have

$$ \theta = \det \bm{u} {\gamma \sigma(v)} { v }{ \sigma(u)}.$$
The first assumption on $\theta$ gives us a contradiction.

Now we consider the case $u=0$.
Comparing entries $(1,1)$ gives us the equality
$$\gamma \sigma(v)^2 = \gamma \sigma(v^2) =\theta.$$

By the second assumption that $\theta \not = \gamma ~(\mod K^{\times 2})$, we obtain a contradiction.

For the converse, first suppose that $\theta = \gamma v^2$ for some element $v \in K$. We show that 
$\theta = z\sigma(z)$, where $z =\bm {0} {\gamma \sigma(v)}{v}{0} $. We have $z\sigma(z) = \bm{\gamma \sigma(v^2)}{0}{0}{\gamma v^2}$. Note that since $\theta, \gamma \in F$, the assumption that $\theta = \gamma v^2$ implies that $v^2 \in F$. Hence $v^2 = \sigma(v^2)$ and we have therefore just shown that $\theta = z\sigma(z)$.

Now suppose $\theta = \det\bm{u} {\gamma \sigma(v)} { v }{ \sigma(u)}$, where $v \in \sqrt{a}F$.
We already demonstrated how $\theta = z\sigma(z)$, where $z = \bm{u} {\gamma \sigma(v)} { v }{\sigma(u)}$.
\end{IEEEproof}

Note that since our construction allows the ground field $F$ to be of arbitrary degree over $\Q$,
it can moreover be used for the multiblock construction in \cite{LMG}.

In the following corollary, we show cases where we can meet the conditions of the previous lemma.
\begin{corollary}\label{cor:criteria}
Let $F$ be a totally real number field, $K = F(\sqrt{a})$, where $a < 0$. Let $\gamma < 0$, and let $\theta < 0$ satisfy $\theta \not = \gamma~(\mod K^{\times 2})$.
Then nonzero elements of $\A$ have nonzero determinant.
\end{corollary}

\begin{IEEEproof}
Note that by our choice of $\gamma<0, a < 0$, the determinant of
$\bm{u} {\gamma \sigma(v)} { v }{ \sigma(u)}$
is positive. Hence it cannot be equal to $\theta$, which is by our choice negative.
Together with the assumption that  $\theta \not = \gamma ~(\mod K^{\times 2})$, both conditions of Lemma \ref{Condition} are satisfied and we conclude that nonzero elements of $\A$ have a nonzero determinant.
\end{IEEEproof}

\begin{example}
In Example \ref{HQ}, $F=\QQ$, $K=\QQ(i)$ and $\gamma=-1$. Thus any choice of $\theta\in F$ with $\theta<0$ not equal to $-1 ~(\mod \QQ^{\times 2})$ will give a division algebra.
\end{example}

%%%%%%%%%%%%%%%%%%%%%%%%%%%%%%%%%%%%%%%%%%%%%%%%%%%%%%%%%%%%%%%%%%%%%%%%%%%%%%%%%%%%%%%%%%%%%%%%%%%%%%%%%%%%%%%%%%%%%
\subsection{Quadratic Forms Criteria for Full Diversity}

We rephrase the conditions of Lemma \ref{Condition} in terms of quadratic forms. We can then use the theory of quadratic forms in order to meet these conditions. First we give some basic definitions.

\begin{defi} Let $F$ be a field. A \emph{quadratic form} on $V$ over $F$ is a map $\vf: V\to F$ satisfying the following two properties:
\begin{enumerate}
\item
$\vf(\alpha \vc v)=\alpha^2 \vf(\vc v)$ for all $\vc v\in V$ and all $\alpha \in F$;
\item the map $b_\vf: V\times V\to F$ defined by
$b_\vf(\vc v,\vc w):=\vf(\vc v+\vc w)-\vf(\vc v)-\vf(\vc w)$
for all $\vc v,\vc w\in V$ is bilinear (and automatically symmetric).
\end{enumerate}
\end{defi}

\begin{defi}
A quadratic form $\vf$ on $V$ is called \emph{anisotropic over $F$} if it has no nontrivial solutions, i.e.,  $\vf(\vc v)=0$ if and only if $\vc v = \vc 0$.
\end{defi}

We will only deal with quadratic forms which have the form $\psi(\vc v)=\sum_{i=1}^n a_i v_i^2$
for all $\vc v=(v_1,\ldots, v_n)\in F^n$ and scalars $a_1,\ldots, a_n\in F$, for convenience employing the short-hand notation $\psi=\<a_1,\ldots, a_n\>.$

The following lemma rephrases Lemma \ref{Condition}.

\begin{lemma}\label{QFcriteria}
Suppose the quadratic form $\langle 1, -a, \gamma a, -\theta \rangle$ is anisotropic over $F$ and $\< \gamma, -\theta \>$ is anisotropic over $K$. Then nonzero elements of $\A$ have nonzero determinant.
\end{lemma}
\begin{IEEEproof}
Since $\< \gamma, - \theta \>$  is anisotropic over $K$, there are no nontrivial solutions to the equation $\gamma x^2 - \theta y^2 =0$, which implies that $\gamma \not = \theta ~(\mod K^{\times 2})$, giving us the second condition of Lemma \ref{Condition}.
Next we verify that the first condition of Lemma \ref{Condition} is satisfied. We will show that for any $u\in K, v \in \sqrt{a}F$, we have
$ \theta \not = u\sigma(u)+\gamma v^2$.
Suppose, for the sake of contradiction, that it is not so. Rewriting with $u = u_0+\sqrt{a}u_1$, $v = \sqrt{a}v_1$ we obtain
$u\sigma(u)+\gamma v^2 = u_0^2 - au_1^2 + \gamma a v_1^2 - \theta = 0$. But this implies that the quadratic form $\langle 1, -a, \gamma a, -\theta \rangle$ has a nontrivial solution in $F$, contrary to our assumption.
\end{IEEEproof}
{ {
The next result is an application of Springer's Theorem. 
\begin{corollary}
Let $\mathcal O_F$ denote the ring of algebraic integers of $F$. Suppose 
 $a$ generates a prime ideal  of $\mathcal O_F$ and that $\theta, \gamma$ are non-square modulo $ a \mathcal O_F$. Then the quadratic form $\langle 1, -a, \gamma a, -\theta \rangle$ is anisotropic.
\end{corollary}
}}

\begin{IEEEproof}
This is an application of Springer's Theorem (Theorem 6.1 \cite{UM}), from which it follows that the quadratic form
$\langle 1, -a, \gamma a, -\theta \rangle$
is anisotropic if and only if the residue forms
$\<1, -\theta \>$, $\<\gamma, -1 \>$
are anisotropic over the residue field $\mathcal{O}_F/{a\mathcal O_F}$. Since $\theta, \gamma$ are both non-squares $(\mod a \mathcal O_F)$, there are no nonzero solutions to the equations $y^2 = \theta z^2 ~(\mod a)$, $y^2 = \gamma z^2 ~(\mod a \mathcal O_F)$. In other words the residue forms $\<1, -\theta \>$ , $\<1, -\gamma\>$ are anisotropic and we obtain the conclusion. For a more detailed version of a nearly identical proof see Theorem 7.1 \cite{UM}.
\end{IEEEproof}

Putting the above together we obtain the following.

\begin{corollary}
Let $a$ generate a prime ideal of $\mathcal O_F$. Suppose $\theta, \gamma$ are both non-square $(\mod a)$ and also suppose $\<\gamma, -\theta \>$ is anisotropic over $K$. Then nonzero elements of $\A$ have a nonzero determinant.
\end{corollary}

\begin{IEEEproof}
By the previous corollary and Lemma \ref{QFcriteria},
the conditions of Lemma \ref{Condition} are satisfied.
\end{IEEEproof}

\begin{example}
Let $F = \Q(i)$ and let $Q = \qa 3 \gamma F$, so that $K=F(\sqrt{3})$. We would like to pick $\gamma, \theta \in F$ which satisfy the criteria of the previous corollary, and hence Lemma \ref{Condition}.
Note that $3\Z[i]$ is a prime ideal of $\Z[i]$ of norm 9, i.e., $\Z[i]/3\Z[i]$ is a finite field $\mathbb F_9$ of $9$ elements.
Let $\gamma = 1+i, \theta = 1-i$. We can directly verify that both $\theta, \gamma$ are non-square modulo $3\Z[i]$ and moreover $\gamma/\theta = i \not \in F(\sqrt{3})^{\times 2}$. Hence, the code resulting from $\alpha_{\theta}(Q,Q)$ is fully diverse.
\end{example}

%***********************************************************************************************************************%
%
% CODES
%
%***********************************************************************************************************************%

\section{Examples of MIDO Code Constructions}
\label{sec:MIDO}

In this section we illustrate the general iterated construction using a few well known codes.

%%%%%%%%%%%%%%%%%%%%%%%%%%%%%%%%%%%%%%%%%%%%%%%%%%%%%%%%%%%%%%%%%%%%%%%%%%%%%%%%%%%%%%%%%%%%%%%%%%%%%%%%%%%%%%%%%%%%%%%%%%%%%%
\subsection{An Iterated MIDO Silver code}
The Silver code, discovered in \cite{HT}, and rediscovered in \cite{PGA},
is given by codewords of the form
\[
\begin{bmatrix}
x_1 & -\cc x_2 \\ x_2 & \cc x_1
\end{bmatrix}+
\begin{bmatrix}
1 & 0 \\ 0 & -1
\end{bmatrix}
\begin{bmatrix}
z_1 & -\cc z_2 \\ z_2 & \cc z_1
\end{bmatrix},~
\]
where
\[
\begin{bmatrix}
z_1 \\ z_2
\end{bmatrix}
=
\frac{1}{\sqrt{7}}
\begin{bmatrix}
1+i & -1+2i \\ 1+2i & 1-i
\end{bmatrix}
\begin{bmatrix}
x_3 \\ x_4
\end{bmatrix}
\]
and $x_1,x_2,x_3,x_4 \in \Z[i]$ are the information symbols.
Consider the number field $F = \Q(\sqrt{-7})$ and its Galois extension $K = F(i)$, with Galois automorphism $\sigma: a+bi \mapsto a-bi$, where $a, b \in F$. Note that $\sigma$ is not the complex conjugation since it fixes $\sqrt{-7}$. It was shown in
\cite{HLRVV} that Silver codewords can alternatively be viewed as (scaled) matrices over $K$ of the form
$$
\begin{bmatrix}
  c & -\sigma(d) \\  d & \sigma( c)
\end{bmatrix},
$$
where $c,d \in \Z[i] \oplus \Z[i](\frac{1+\sqrt{-7}}{2}).$ To be more specific, let $Q = \qa {-1} {-1} F$ be a quaternion algebra with $F = \Q(\sqrt{-7})$. The Silver code
is obtained via { {matrix representation}} of the elements of the order $\mathcal{O}_K \oplus j \mathcal{O}_K$ of $Q$, that is, for each element
$c+jd$, with $c,d \in \Q(\sqrt{-7},i)$ we have
$$c+jd \mapsto
\begin{bmatrix}
  c & -\sigma(d) \\  d & \sigma( c)
\end{bmatrix}.
$$

Now for any  $\theta\in F$, we have the following mapping, which sends two elements of the Silver code to an element of the iterated Silver code:
\begin{equation}\label{eq:silveri}
\alpha_{\theta} : \left(
\begin{bmatrix}
  c & -\sigma(d) \\  d & \sigma(c)
\end{bmatrix},
\begin{bmatrix}
  e & -\sigma(f) \\  f & \sigma( e),
\end{bmatrix}
\right)
\mapsto  
\end{equation}
\[
\begin{bmatrix}
c & - \sigma(d) &  \theta\sigma(e) & -\theta f \\
d & \sigma(c)   &  \theta\sigma(f) & \theta e \\
e & - \sigma(f) &  \sigma(c) & -  d\\
f & \sigma(e)   &  \sigma(d) &    c \\
\end{bmatrix}.
\]
When $\theta$ satisfies conditions 1), 2) of Lemma \ref{Condition}, the resulting code will be fully diverse. Next we show this is the case for  $\theta = -17$.

Clearly, $\theta/{-1} = 17$ is  not a square in $K$.

{ {
Condition 1) amounts to showing that
$\theta \neq u\sigma(u) - \gamma v\sigma(v)$, where $u \in K, v \in iF$. 
Writing $u= u_0 +i u_1, v = i v_1$ where $u_0, u_1, v_1 \in F$, and recalling that 
$\sigma: i \mapsto -i$ and $\gamma = -1$, we thus want to show that 
$\theta \neq u^2_0+u^2_1+v^2_1$ 
for any $u_0, u_1, v_1 \in F$. 
We show that $\theta$ is not a sum of three squares by considering reduction modulo the ideal $I=({\frac{1+\sqrt{-7}}{2}})^3\mathcal O_F$. First note that $\mathcal O_F/({\frac{1+\sqrt{-7}}{2}}) \cong\Z/2\Z$,  $\mathcal O_F/I^3 \cong \Z/8\Z$ and observe that $\theta \equiv -1 ~(\mod I)$. Using the isomorphism $\mathcal O_F/I^3 \cong \Z/8\Z$, we verify that $-1~(\mod I)$ is not a sum of three squares, and hence neither is $\theta$. \\
We conclude that the iterated Silver code is fully diverse, in fact it has NVD since $F$ is imaginary quadratic. 
}}
 
As an example of a basis of an iterated MIDO Silver code, fix the natural basis of the Silver code, and consider its image under $\alpha_{\theta}$ :
%{\tiny{
\begin{equation}\label{SilverBasis}
\alpha_{\theta}\left( \bm 1 0 0 1,0\right),~
\alpha_{\theta}\left(\bm i 0 0 {-i},0\right),~\end{equation}
\[
\alpha_{\theta}\left(\bm 0 {-1} 1 0,0\right),~
\alpha_{\theta}\left(\bm 0 i i 0,0\right),~
\]
\[
\alpha_{\theta}\left(\frac{1}{\sqrt{7}}\bm  {  1 + i}  {-1 + 2i} {  -1 - 2i} {  -1 + i},0\right),~\]
\[
\alpha_{\theta}\left(\frac{1}{\sqrt{7}}\bm   {-1 + i}  { 2 + i}  { 2 - i}   {1 + i},0\right),~
\]
\[
\alpha_{\theta}\left(\frac{1}{\sqrt{7}}\bm   {-1 + 2i}  {-1 - i}  {-1 + i}   {1 + 2i},0\right),~ \]
\[
\alpha_{\theta}\left(\frac{1}{\sqrt{7}}\bm   {-2 - i}  {-1 + i}  {-1 - i}   {2 - i},0\right),~
\]
\[
\alpha_{\theta}\left(0, \bm 1 0 0 1\right),~
\alpha_{\theta}\left(0,\bm i 0 0 {-i}\right),~
\alpha_{\theta}\left(0,\bm 0 {-1} 1 0\right),~
\]
\[
\alpha_{\theta}\left(0,\bm 0 i i 0\right),~
\alpha_{\theta}\left(0,\frac{1}{\sqrt{7}}\bm  {  1 + i}  {-1 + 2i} {  -1 - 2i} {  -1 + i}\right),~
\]
\[
\alpha_{\theta}\left(0,\frac{1}{\sqrt{7}}\bm   {-1 + i}  { 2 + i}  { 2 - i}   {1 + i}\right),~\]
\[
\alpha_{\theta}\left(0,\frac{1}{\sqrt{7}}\bm   {-1 + 2i}  {-1 - i}  {-1 + i}   {1 + 2i}\right),~\]
\[
\alpha_{\theta}\left(0,\frac{1}{\sqrt{7}}\bm   {-2 - i}  {-1 + i}  {-1 - i}   {2 - i}\right)~ .
\]

We start with a general lemma which holds for any choice of $\theta$. We use the iterated MIDO Silver code as an illustration. However, the same result will hold for iterated codes obtained via $\alpha_\theta$ from $Q = \qa a {-1} F$, where $a < 0$ and $F$ is a quadratic extension of $\Q$. In the case of the Silver code $F = \Q(\sqrt{-7})$ and $a = -1$.

\begin{lemma}\label{ImQuadratic}
The complexity of an iterated MIDO Silver code is at most $O(|S|^{13})$, no matter the choice of $\theta$.
\end{lemma}
\begin{IEEEproof}
We may choose a basis for the Silver code so that it contains the matrices
\[
\bm 1 0 0 1,~\bm i 0 0 {-i},~\bm 0 {-1} 1 0,~\bm 0 i i 0
\]
so that
$$
B_1=\alpha_{\theta}\left(\bm 1 0 0 1, 0\right),
B_2=\alpha_{\theta}\left(\bm i 0 0 {-i}, 0\right),
$$
$$
B_3= \alpha_{\theta}\left(\bm 0 {-1} 1 0, 0\right),
B_4 = \alpha_{\theta}\left(\bm 0 i i 0, 0\right)
$$
are four basis matrices for an iterated MIDO Silver code, independently of $\theta$.
It is a direct computation to check that
$$
B_kB_l^* + B_lB_k^*=0,\mbox{for all }k \not = l \in \{1, \ldots, 4\}.
$$

By choosing an ordering on the basis of an iterated MIDO Silver code such that the first four elements are
$B_1,B_2,B_3,B_4$, the above computations show that the associated matrix $M$, defined in (\ref{orthRelations}), has the structure 
\begin{equation}\label{eq:M13}
M = \bm {\Delta} {M_1} {M_2} {M_3},
\end{equation}
where $\Delta$ is a $4\times 4$ diagonal matrix. 
Hence, by \cite[Lemma 2]{JR}, the decoding complexity reduces from $O(|S|^{16})$ to $O(|S|^{13})$.
\end{IEEEproof}

When $\theta = \zeta \theta'$ for $\zeta^4=1$ we get a better reduction in decoding complexity using the map $\tilde\alpha:=\tilde \alpha_{\zeta\sqrt{\theta}}$ as defined in (\ref{eq:alphatilde}): 
$$ 
\tilde\alpha : \left(
\begin{bmatrix}
  c & -\sigma(d) \\  d & \sigma(c),
\end{bmatrix} ,
\begin{bmatrix}
  e & -\sigma(f) \\  f & \sigma( e),
\end{bmatrix}
\right)
$$
\begin{equation}\label{eq:silver1}
\mapsto
\begin{bmatrix}
c & - \sigma(d) &  \zeta \sqrt{\theta'}\sigma(e) & -\zeta\sqrt{\theta'} f \\
d & \sigma(c)   &  \zeta \sqrt{\theta'}\sigma(f) & \zeta\sqrt{\theta'} e \\
\sqrt{\theta'}e & - \sqrt{\theta'}\sigma(f) &  \sigma(c) & -  d\\
\sqrt{\theta'}f & \sqrt{\theta'}\sigma(e)   &  \sigma(d) &    c \\
\end{bmatrix}.
\end{equation}

We demonstrate this further reduction in complexity in the following lemma.

\begin{lemma}\label{ImQuadraticSilver}
The complexity of the iterated MIDO Silver code (\ref{eq:silver1}) with  $\theta=- \theta'$ is $O(|S|^{10})$.
\end{lemma}
\begin{IEEEproof}
We show that in fact the code is conditionally $4$-group decodable.
Using the basis for the Silver Code given in (\ref{SilverBasis}), in this exact order, define the partitions
\[
\Gamma_1=\{1,11\},\Gamma_2=\{3,9\},\Gamma_{3}=\{4,10\},\Gamma_{4}=\{2,12\},
\]
\[
\Gamma^C=\{ 5,    6,    7,    8,   13,   14,  15,  16\}.
\]
It can be verified by direct computation that the iterated MIDO Silver code with $\theta=-\theta'$ has   matrix $M$, whose first $8$ rows have the form
$$
{
 \left[
{\begin{array}{cccccccc}
t & 0 & 0 & 0 & t & t & t & t \\
0 & t & 0 & 0 & t & t & t & t \\
0 & 0 & t & 0 & t & t & t & t \\
0 & 0 & 0 & t & t & t & t & t \\
\end{array}}
 \right]}
$$
where $0$ denotes a $2\times 2$ matrix and $t$ any $2\times 2$ nonzero matrix. It can thus be seen
that the iterated Silver code is conditionally $4$-group decodable: the four groups of two symbols are
clearly distinct.  The meaning of {\emph{conditionally group decodable}} is also  readily understood:  conditioned
on decoding the  last $8$ information symbols (for a cost of $O(|S|^{8})$), the other half of the symbols is
decodable by pair (for a cost of $O(|S|^{2})$) independently,
resulting in a complexity of $O(|S|^{10})$.
\end{IEEEproof}

Furthermore, for the sake of experiment, we tried the choice of $\theta=i$, another root of unity which seems very natural, even though it belongs to $K$ and thus is not included in the description of our iterated code construction.
An explicit computation of the $M$ matrix when $\theta=i$ shows that it is of the form (\ref{eq:M13}).
Consequently, the decoding complexity of the code (\ref{eq:silveri}) when
$\theta=i$ is $O(|S|^{13})$.

\begin{figure}
\centering
\includegraphics[width=3.2in]{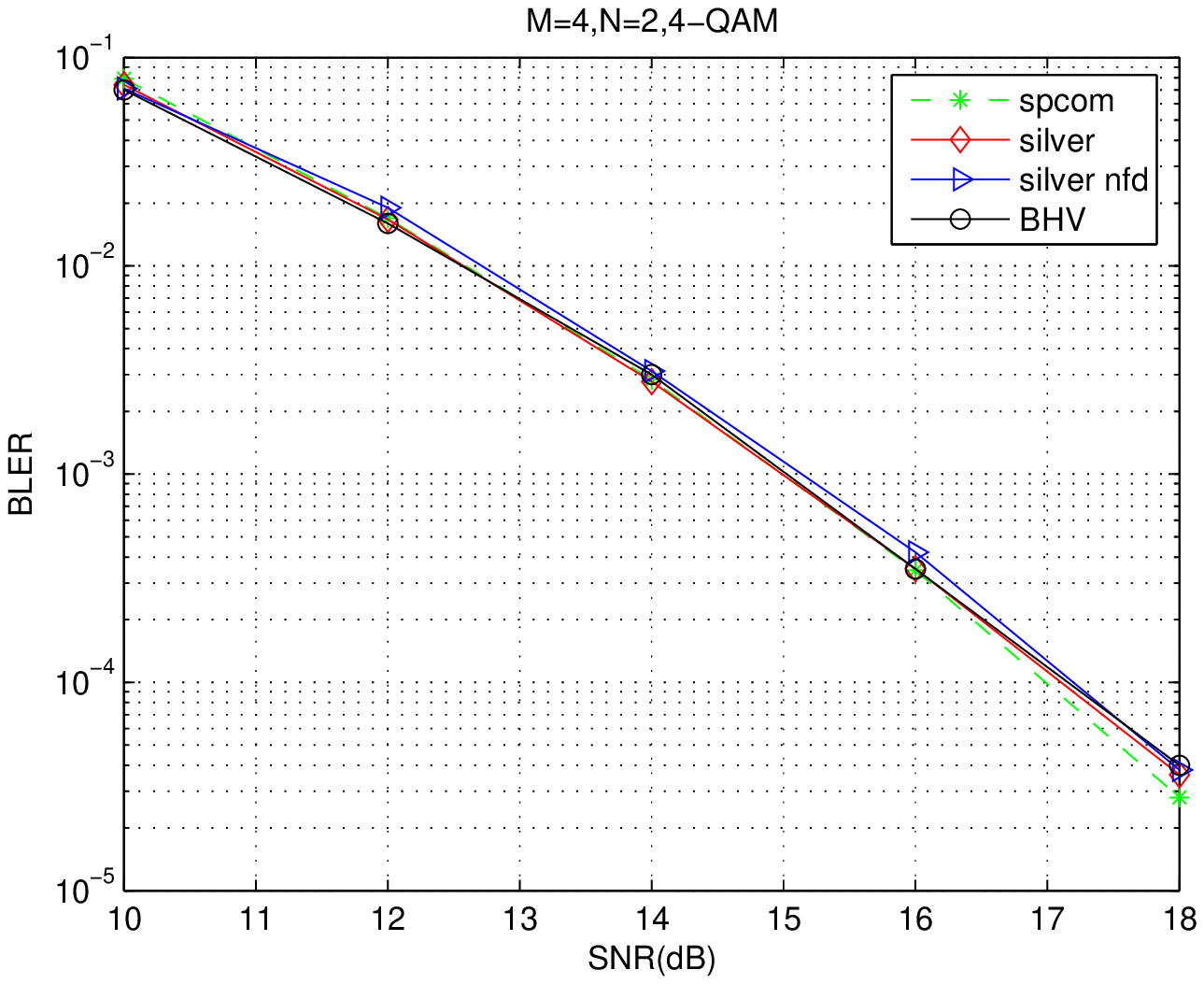}
\includegraphics[width=3.2in]{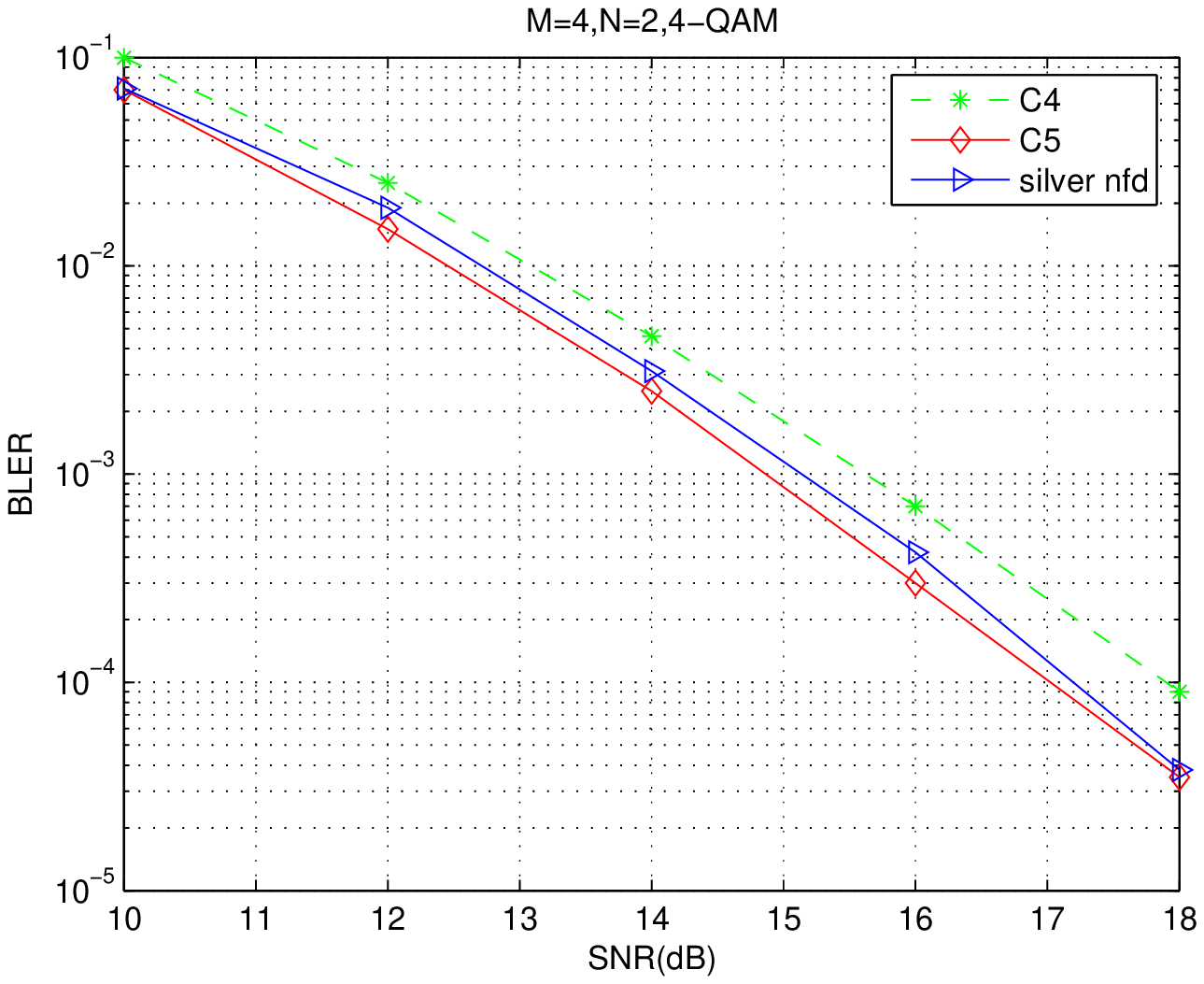}
\caption{Comparison among codes with decoding complexity $O(|S|^{12})$ and $O(|S|^{13})$ on the left, 
and of decoding complexity $O(|S|^{10})$ on the right.}
\label{fig:silver}
\end{figure}

The performance of two versions of the iterated MIDO silver code, one with $\theta=-1$ of complexity $O(|S|^{10})$ and the other one with $\theta=i$ and complexity $O(|S|^{13})$, is illustrated on
Fig.~\ref{fig:silver}. On both figures, the $x$-axis corresponds to
the signal-to-noise ration (SNR) in dB, and the $y$-axis to the Block Error Rate (BLER).
The signal constellation $S$ used is a 4-QAM constellation. On the left of Fig.~\ref{fig:silver}, codes
with decoding complexity $O(|S|^{12})$ and $O(|S|^{13})$ are displayed: we see that the difference among the codes is tiny up to 18 dB. The BHV code \cite{BHV} has complexity $O(|S|^{12})$ and is known not to be fully diverse. It has a very good performance, and this is only starting from 18 dB that the loss in diversity starts to be noticed. The SPCOM code \cite{OHV} has worse decoding complexity than the BHV code, but is fully diverse, and thus gives a way of comparing the gain in diversity with respect to the BHV code, which is best seen at higher SNR. The iterated MIDO Silver code with $\theta= i$ also starts to exhibit a loss in diversity around 18 dB. The iterated MIDO Silver code with $\theta=-1$, { {which is not fully-diverse,}} is also shown for the sake of comparison. The fact that this code is not fully diverse can be seen much earlier than for the BHV code, though its final performance in this range of SNR is quite similar. To get a fairer comparison, the iterated MIDO Silver code which has decoding complexity $O(|S|^{10})$ is compared with two codes of same complexity, based on crossed product algebras \cite{LO}. The code $C_5$ is not fully diverse, but thanks to a good shaping, has very good performance from low to medium SNR. The code $C_4$ is fully diverse, but its performance suffers from the lack of cubic shaping. The newly proposed code behaves closely to the code $C_5$.

\subsection{An Iterated MIDO Golden Code} \label{ex:GC}
Consider the quaternion algebra $Q = \qa {5} {i} {\Q(i)}$, with $K=\Q(i,\sqrt{5})$, $\sigma: \sqrt{5} \mapsto -\sqrt{5}$.
Let $\phi=(1+\sqrt{5})/2$ denote the Golden number, with $\sigma(\phi)=(1-\sqrt{5})/2$.
Set $\beta =1+i\sigma(\phi)$, with $\sigma(\beta )=1+i\phi$.
We can view elements of the order $\{ \beta  q,~q\in Q\}$ of $Q$ as $2\times 2$ matrices over $K$ via { { representation in $M_{n\times n}(K)$}} as
$$
\begin{bmatrix}
\beta  c & i \sigma(\beta  d) \\ \beta  d & \sigma(\beta  c)
\end{bmatrix},
$$
which, up to a scaling by $1/\sqrt{5}$, are codewords of the Golden code \cite{BRV, DV}.
Now define, for a suitable $\theta \in \QQ(i)$, the map 
$$ 
\alpha_{\theta} : 
\left(
\begin{bmatrix}
\beta  a & i \sigma(\beta  b) \\ \beta  b & \sigma(\beta  a)
\end{bmatrix},
\begin{bmatrix}
\beta  c & i \sigma(\beta  d) \\ \beta  d & \sigma(\beta  c)
\end{bmatrix}
\right)
\mapsto
$$
$$
\begin{bmatrix}
\beta  a & i \tau(\beta )\tau(b) &  \theta\tau(\beta )\tau(c) & i\theta \beta  d \\
\beta  b & \tau(\beta )\tau(a)   &  \theta\tau(\beta )\tau(d) & \theta \beta  c \\
\beta  c & i \tau(\beta )\tau(d) &  \tau(\beta )\tau(a) & i\beta  b\\
\beta  d & \tau(\beta )\tau(c)   &  \tau(\beta )\tau(b) &  \beta  a \\
\end{bmatrix}.
$$

{ {
By Lemma \ref{normConditionN}, the determinant of each codeword is in $\Q(i)$, moreover it is indeed in $\Z[i]$, since the coefficients of each codeword come from the ring of algebraic integers of $K$. 
Hence in order to show that the iterated Golden code has NVD, it suffices to pick suitable $\theta$, so that the determinant of each nonzero codeword is nonzero. In the following lemma we do so by applying Lemma \ref{QFcriteria}. 
}}
\begin{lemma}
Let $\theta = 1- i$. Then the iterated Golden code arising from $\alpha_\theta(Q,Q)$ is fully-diverse and has NVD.  
\end{lemma}
\begin{IEEEproof}

Lemma \ref{QFcriteria} says that it suffices 
to pick $\theta \in \Q(i)$ so that the quadratic form $\<1, -5, 5\gamma, -\theta\>$ is anisotropic over $ \Q(i)$, and $\< \gamma, -\theta \>$ is anisotropic over $K$.
Recalling that $\gamma = i$, we need to show that $\<1,-5, 5i, -(1-i)\>$ is anisotropic. Note that $5$ factors into $5=(2+i)(2-i)$ in $\Q(i)$, each of the factors generating a prime ideal. By Springer's Theorem \cite[Theorem 6.1]{UM}, it suffices to show that the quadratic forms $\<1, -\theta\>$ and $\<-(2-i), (2-i)i\>$ are anisotropic over the finite field $\Z[i]/(2+i)$. Showing that $\<1, -\theta\>$ is anisotropic is equivalent to establishing that $\theta$ is not a square in $\Z[i]/(2+i)\cong \mathbb F_5$. By our choice  $\theta = 1-i \equiv 3 ~\mod {(2+i)}$, which is not in $\mathbb F_5^{\times 2} = \{\pm 1\}$. 
Similarly showing that $\<-(2-i), (2-i)i\>$ is anisotropic over $\Z[i]/(2+i)$ is equivalent to establishing that $i$ is not a square in $\Z[i]/(2+i)$, which follows from the fact that $8$ does not divide the order of $\F_5^\times$.

It remains to show that $\theta/\gamma \not\in K^{\times 2}$. Let $z = \theta/\gamma = \frac{1-i}{i} = -i-1$. Suppose $z = x^2$. Then $zz^* =(-i-1)(i-1) = 2 = (xx^*)^2$. However, clearly $\sqrt{2} \not \in K$, which completes the argument.  

\end{IEEEproof}

%%%%%%%%%%%%%%%%%%%%%%%%%%%%%%%%%%%%%%%%%%%%%%%%%%%%%%%%%%%%%%%%%%%%%%%%%%%%%%%%%%%%%%%%%%%%%%%%%%%%%%%%%%%%%%%%%%
\subsection{An Iterated MIDO Alamouti Code}\label{ex:Alamouti}

We consider the iterated construction of Example \ref{HQ}, with $Q = \qa {-1} {-1} \Q$, $F=\QQ$, $K=\QQ(i)$ and for $x\in\QQ(i)$, 
$\sigma:x\mapsto x^*$ denotes the complex conjugation. We view elements of $Q$ as $2\times 2$ matrices over $K=\Q(i)$ of the form (\ref{eq:alamcodeword}), corresponding to the celebrated Alamouti code \cite{Alamouti}. Now
$$
\alpha_{\theta} : 
\left(
\begin{bmatrix}
f & -\cc{g} \\
g & \cc{f}\\
\end{bmatrix},
\begin{bmatrix}
c & -\cc{d} \\
d & \cc{c}\\
\end{bmatrix}
\right)
\mapsto
\begin{bmatrix}
f & - \cc{g} &  \theta\cc{c} & -\theta d \\
g & \cc{f}   &  \theta\cc{d} & \theta c \\
c & - \cc{d} &  \cc{f} & - g\\
d & \cc{c}   &  \cc{g} &  f \\
\end{bmatrix}
.
$$
The case $\theta=-1$ corresponds to the quasi-orthogonal code proposed by Jafarkhani
\cite{Jafarkhani}. However, this code is not fully diverse, as already noticed in
\cite{Jafarkhani}. This can be remedied by replacing $-1$ with any negative $\theta$ to satisfy the criteria of Lemma \ref{Condition}, as long as $\theta/\gamma = -\theta \not \in \Q(i)^{\times 2}$, e.g. $\theta = -3$.
However this code is not full-rate, as it can transmit up to $4$ complex, equivalently $8$ real information symbols.

In order to increase the rate of such a code, we could instead try to consider the quaternion algebra
$Q=(-1,-1)_{\QQ(\sqrt{b})}$, or more generally $Q=(-a,-\gamma)_{\QQ(\sqrt{b})}$, with
$a,b,\gamma>0$, as proposed in \cite{MO11}. This in fact does give very good orthogonality relations among the basis elements, as shown 
in \cite{MO11}. However, choosing as a base field a totally real quadratic number field means that the basis matrices $\{D_1, \ldots, D_8\}$ generating $Q$ over $\QQ$ will be linearly dependent. To see that, recall that $\{D_1, \ldots, D_8\}$ are 
\[
D_1=
\begin{bmatrix}
1 & 0 \\ 0 & 1
\end{bmatrix},
D_2=
\begin{bmatrix}
\sqrt{b} & 0  \\ 0 & \sqrt{b}
\end{bmatrix},
D_3=
\begin{bmatrix}
\sqrt{-a} & 0 \\ 0 & -\sqrt{-a}
\end{bmatrix},
\]
\[
D_4=
\begin{bmatrix}
\sqrt{-a}\sqrt{b} & 0 \\ 0 & -\sqrt{-a}\sqrt{b},
\end{bmatrix},
D_5=
\begin{bmatrix}
0 & -\sqrt{\gamma} \\ \sqrt{\gamma} & 0
\end{bmatrix},
\]
\[
D_6=
\begin{bmatrix}
0 &-\sqrt{\gamma}\sqrt{b} \\  \sqrt{\gamma}\sqrt{b} & 0
\end{bmatrix},
D_7=
\begin{bmatrix}
0 & \sqrt{\gamma}\sqrt{-a}  \\ \sqrt{\gamma} \sqrt{-a} & 0
\end{bmatrix},
\]
\[
D_8=
\begin{bmatrix}
0 & \sqrt{\gamma}\sqrt{-a}\sqrt{b}  \\ \sqrt{\gamma}\sqrt{-a}\sqrt{b} & 0
\end{bmatrix}
.
\]
We can observe that $D_1$ and $D_2$ are real multiples of each other, implying that 
\[
B_1=\alpha_{\theta}(D_1,0)=
\begin{bmatrix}
I_2 & 0 \\
0  & I_2
\end{bmatrix},~
B_2=\alpha_{\theta}(D_2,0)=
\sqrt{b}
\begin{bmatrix}
I_2 & 0 \\
0  & I_2
\end{bmatrix}
\]
are real multiples of each other as well. This affects the rank of the matrix $B$ in (\ref{matrixB}), thus 
that of $R$; more specifically, the matrix $R$ is no longer full-rank. As a consequence, we cannot draw conclusion about fast-decodability of the code, despite the sparsity of $M$.

This can be avoided by using a totally imaginary quadratic field as base field. The complexity induced in such a case is similar 
to that of the Silver code, discussed above.

%***********************************************************************************************************%
%
% CODES NEXT

%***********************************************************************************************************%

\section{Examples of Higher Dimensional Code Constructions}
\label{sec:higher}

We now explain how our general construction applies to the case of an asymmetric MIMO channel with $6$ transmit and $3$ receive antennas \cite{MOspcom}.

%*************************************************************************************%
\subsection{Cyclic Algebras of Degree 3}

\begin{figure}
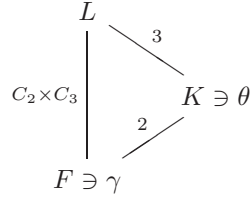

\[
\begin{diagram}
\node{L}\arrow{se,l,-}{3}\arrow[2]{s,l,-}{C_2\times C_3}\\
\node{}\node{K\ni\theta}\arrow{sw,l,-}{2}\\
\node{F\ni\gamma}
\end{diagram}
\]
\caption{
\label{fig:ext}
Field extension used for the proposed iterative code construction.
}
\end{figure}

Let $L/F$ be a Galois extension of number fields with Galois group $C_2\times C_3\simeq C_6$, where
$\sigma$ denotes the generator of the cyclic group $C_3$ and $\tau$ that of $C_2$ (see Fig.~\ref{fig:ext}).
Let $K$ be the fixed field of $\sigma$, so that $L/K$ forms a cyclic extension of degree $3$. 
Let $\Dc$ be the corresponding cyclic algebra of degree 3. 
The { {representation $\lambda$}} of elements of $\Dc$ is given by
\begin{equation}\label{eq:code3x3}
\begin{bmatrix}
a & \gamma \sigma(c) & \gamma\sigma^2(b) \\
b & \sigma(a) & \gamma\sigma^2(c) \\
c & \sigma(b) & \sigma^2(a)\\
\end{bmatrix},~a,b,c\in L,
\end{equation}

where $\gamma\in K$ is chosen so that $\Dc$ is division, which guarantees full diversity (\ref{eq:fulldiv}).

Given an element $\theta$ of $K$, let  $\alpha_{\theta}: \Dc\times \Dc \rightarrow M_2(\Dc)$ be the map defined in (\ref{eq:alpha}) by
$$
\alpha_{\theta} : (x,y) \mapsto
\begin{bmatrix}
  x & \theta\tau(y) \\  y & \tau(x)
\end{bmatrix},
$$
where $x,y$ are identified with their { {representation $\lambda(x), \lambda(y)$,}} and $\tau(x)$ is defined componentwise, so that
{ 
{
$$
\alpha_{\theta} : \left(
\begin{bmatrix}
a & \gamma \sigma(c) & \gamma\sigma^2(b) \\
b & \sigma(a) & \gamma\sigma^2(c) \\
c & \sigma(b) & \sigma^2(a)\\
\end{bmatrix},
\begin{bmatrix}
a' & \gamma \sigma(c') & \gamma\sigma^2(b') \\
b' & \sigma(a') & \gamma\sigma^2(c') \\
c' & \sigma(b') & \sigma^2(a')\\
\end{bmatrix}
\right)\mapsto
$$
\begin{equation}\label{eq:codeword}
\begin{bmatrix}
a & \gamma \sigma(c) & \gamma\sigma^2(b)& 
\theta\tau(a') & \theta\gamma \tau\sigma(c') & \theta\gamma\tau\sigma^2(b') \\
b & \sigma(a) & \gamma\sigma^2(c)&
 \theta\tau(b') & \theta\tau\sigma(a') & \theta\gamma\tau\sigma^2(c') \\
c & \sigma(b) & \sigma^2(a)& 
\theta\tau(c') & \theta\tau\sigma(b') & \theta\tau\sigma^2(a')\\
a' & \gamma \sigma(c') & \gamma\sigma^2(b')&
 \tau(a) & \gamma \tau\sigma(c) & \gamma\tau\sigma^2(b) \\
b' & \sigma(a') & \gamma\sigma^2(c') &
 \tau(b) & \tau\sigma(a) & \gamma\tau\sigma^2(c)\\
c' & \sigma(b') & \sigma^2(a') &
 \tau(c) & \tau\sigma(b) & \tau\sigma^2(a)\\
\end{bmatrix}.
\end{equation}
}}
We now describe one scenario when the image $\Ac$ of $\alpha_{\theta}(\Dc,\Dc)$ satisfies the full diversity property.

\begin{lemma}\label{ConditionFD3}
Suppose that $\gamma$ is chosen in $F$, so that $\tau(\gamma)=\gamma$.
Let $\theta$ be an element of $K$ such that $\tau(\theta^3)\neq\theta^3$.
Then $\Ac$ satisfies the full diversity property.
\end{lemma}
\begin{IEEEproof}

By Lemma \ref{normConditionN}, the claim  is equivalent to showing that

$$
XY^{-1}\tau(XY^{-1}) \neq \theta I_3.
$$
Let $Z=XY^{-1}$ and write it as
{ 
{
$$Z =
\begin{bmatrix}
u & \gamma \sigma(w) & \gamma\sigma^2(v) \\
v & \sigma(u) & \gamma\sigma^2(w) \\
w & \sigma(v) & \sigma^2(u)\\
\end{bmatrix},~u,v,w\in L.
$$
}}
For the sake of contradiction, suppose $ Z \tau(Z) = \theta I_3$.
Then
\[
\det(Z)\det(\tau(Z))=\det(Z)\tau(\det(Z)) = \theta ^3.
\]
Since $\det(Z)\tau(\det(Z))$ is fixed by $\tau$, it should be that $\theta^3$ is also fixed by $\tau$,
a contradiction.
\end{IEEEproof}

\begin{remark}
By choosing $\theta$ not in $F$, $\Ac$ is not multiplicatively closed,
thus it does not form an algebra (see Lemma \ref{AisAlgebra}).
\end{remark}

We now provide examples of explicit code constructions that fit the above setting.

%*************************************************************************************%
\subsection{Examples}
{ {
\begin{example}
$\Dc = (L/K, \sigma_2, 1+i)$,  with 
$L = \Q(\zeta_7, i)$, where $\zeta_7$ is a primitive $7^{th}$ root of unity,  
$\sigma_2:\zeta_7 \mapsto \zeta_7^2$
and $K$ is the fixed field $ L^{\<\sigma_2\>}$. 
To define the map $\alpha_\theta$, we consider $\tau$ to be the generator of $Gal(\Q(\sqrt{-7},i)/\Q(i)$ and values 
$\theta = \sqrt{-7}$, $\theta = -1$. 
\label{ex:6x3Example1}
\end{example}
}}

\begin{figure}
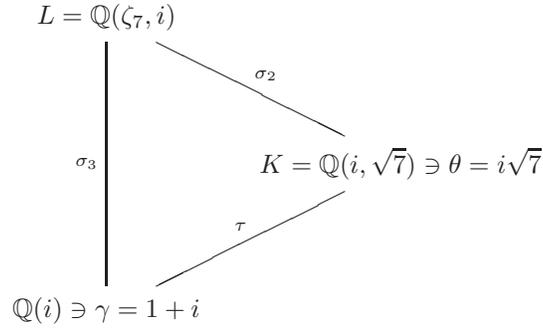

\[
\begin{diagram}
\node{L=\Q(\zeta_7,i)}\arrow{se,l,-}{\sigma_2}\arrow[2]{s,l,-}{\sigma_3}\\
\node{}\node{K=\Q(i,\sqrt{7})\ni\theta=i\sqrt{7}}\arrow{sw,l,-}{\tau}\\
\node{\Q(i)\ni\gamma=1+i}
\end{diagram}
\]
\caption{
\label{fig:extcode1}
Field extension used for the first code example.
}
\end{figure}
The Galois group $Gal(L/\Q(i))$ is cyclic of order $6$, generated by the automorphism
$$\sigma_3:\zeta_7 \mapsto \zeta_7^3.$$
Note that the automorphism
$$\sigma_2:\zeta_7 \mapsto \zeta_7^2$$
is of order $3$ and hence has a fixed field $K$ which is of degree $2$ over $\Q(i)$.
Therefore $K=\Q(\sqrt{-7},i) = \Q(\sqrt{7}, i)$, with the Galois $\Q(i)$-automorphism
$$ \tau: \sqrt{7}\mapsto -\sqrt{7},$$
$$\tau: i \mapsto i.$$

{ 
{
\begin{claim}
The algebra $\Dc = (L/K, \sigma_2, 1+i)$ is division. 
\end{claim}
}}
\begin{IEEEproof}
Note that the ideal $7\Z$ remains inert in $\Z[i]$, in other words, its residue field $\Z[i]/7\Z[i]$ is the finite field $\F_{49}$ of $49$ elements.
We further note that the ideal $7\Z[i]$ is completely ramified in $\Q(\zeta_7, i)$, hence its residue field remains to be of size $49$ in $K$.
By local class field theory, the reciprocity map induces an epimorphism 
{ {
$\F^\times_{49} \rightarrow C_3 \cong Gal(L/K),$
whose kernel is a subgroup of $\F^\times_{49}$ of order 16, which contains the image of norms of $L/K$ (see \cite{LMG} for more detailed treatment of this type of argument).
Hence if an element is a norm of $\Q(\zeta_7, i)/ \Q(i, \sqrt{7})$, then its image in $\F^\times_{49}$ is an element whose order divides $16$. %In particular any element which multiplicatively generates the nonzero elements of $\F_{49}$ cannot be a norm in the extension  $\Q(\zeta_7, i)/ \Q(i, \sqrt{7})$ 
\\
This allows us to conclude that $\gamma = 1+i$, which has order $24$ in $\F^\times_{49}$ is not a norm in the extension $L/K$,
hence the cyclic algebra $(L/K, \sigma_2, \gamma)$ is division. 
This completes the claim. 
}}
\end{IEEEproof}

The choice $\gamma = 1+i$ satisfies the condition  $\tau(\gamma)=\gamma$, as required by the iterated construction.
Note that $\theta=i\theta'=i\sqrt{7}$ satisfies the condition $\tau(\theta^3)\neq \theta^3$ of Lemma \ref{ConditionFD3}. 
{ {
Since $K^{\<\tau\>} = \mathbb Q(i)$,  Lemma \ref{normConditionN} implies that the resulting code has NVD. 
From Lemma \ref{lem:fd}, the iterated code obtained via $\tilde{\alpha}_{i\sqrt{\theta}}(\Dc,\Dc)$ defined in (\ref{eq:alphatilde}) will hence inherit fast-decodable properties of $\Dc$. 
}}
{ 
{
Next we fix a convenient basis for the code arising from $\Dc$. This code should be of $\Q$-rank 18, so that the iterated code is of desired $\Q$-rank 36. 
}}
Let $\{\nu_1,\nu_2,\nu_3\}$ be a $K$-basis of $L$, such that all $\nu_i \in \mathbb{R}$, e.g. $\{1,\zeta_7+\zeta_7^{-1},\zeta_7^2+\zeta_7^{-2} \}$.
The { {representation $\lambda$}} of $\Dc$ gives a family of $3\times3$ matrices of the form
$$\sum_{k=0}^2
\begin{bmatrix}
c_k & 0 & 0 \\
0   & \sigma(c_k) & 0 \\
0 & 0 & \sigma^2(c_k)
\end{bmatrix}
\Gamma^k,$$
where
\[
\Gamma=
\begin{bmatrix}
0 & 0 & \gamma \\
1   & 0 & 0 \\
0 & 0 & 1
\end{bmatrix}
\]
and $c_0, c_1, c_2$ are $K$-linear combinations of $\{\nu_1,\nu_2,\nu_3\}$, that is
$$\sum_{k=0}^2
\sum_{j=1}^3
c_{kj}
\begin{bmatrix}
\nu_j & 0 & 0 \\
0   & \sigma(\nu_j) & 0 \\
0 & 0 & \sigma^2(\nu_j)
\end{bmatrix}
\Gamma^k.$$
We can thus encode $9$ elements of $K$. Since $K$ is of degree 2 over $\QQ(i)$, this is more rate than we can support,
namely $18$ complex symbols instead of $9$. 
{ {Let us write $d\in K$ as
\[
d=d_0+d_1i+\sqrt{7}d_2+\sqrt{7}id_3,~d_1,d_2,d_3,d_4\in\QQ.
\]
By setting two $d_i$ to zero, we will obtain the right rate. Instead of the $\QQ$-basis $\{1,i, \sqrt 7, i\sqrt 7 \}$ of $K$, we will  thus use a subset $\{\mu_1, \mu_2\}$. For reasons of fast-decodability we insist that $\mu_1$ be  real ($\mu_1$ is 1 or $\sqrt{7}$) and $\mu_2$ be totally imaginary ($\mu_2$ is $i$ or $i\sqrt{7}$). 
We obtain from a $K$-basis of $\Dc$ the following $\Q$-linearly independent set for $j=1,2,3$: 
}}
\[
D_j=\mu_1V_j,
~D_{3+j}=
\mu_2
V_j,
\]
\[
D_{6+j}=
\mu_1 V_j\Gamma,~D_{9+j}=\mu_2 V_j\Gamma,
\]
\[
D_{12+j}=
\mu_1
V_j
\Gamma^2,~
D_{15+j}=
\mu_2
V_j\Gamma^2
\]
where we denote
\[
V_j=
\begin{bmatrix}
\nu_j & 0 & 0 \\
0   & \sigma(\nu_j) & 0 \\
0 & 0 & \sigma^2(\nu_j)
\end{bmatrix}
\]
for short.
We have
\[
\mu_1V_j\mu_2^*V_k^*+\mu_2V_k\mu_1^*V_j^*=\mu_1(\mu_2^*+\mu_2)V_jV_k=0
\]
using the fact that $V_j,V_k$ have only real coefficients, and commute since they are diagonal.
Furthermore, $\mu_1$ was also real, while $\mu_2$ is totally imaginary.
Next
\begin{eqnarray*}
&&\mu_1V_j\Gamma\mu_2^*\Gamma^*V_k^*+\mu_2V_k\Gamma\mu_1^*\Gamma^*V_j^*\\
&=& \mu_1\mu_2^* V_j\Gamma\Gamma^*V_k+\mu_2\mu_1V_k\Gamma\Gamma^*V_j^*\\
&=&\mu_1(\mu_2^*+\mu_2)V_j\Gamma\Gamma^*V_k=0
\end{eqnarray*}
since
\[
\Gamma\Gamma^*=
\begin{bmatrix}
0 & 0 & \gamma \\
1   & 0 & 0 \\
0 & 0 & 1
\end{bmatrix}
\begin{bmatrix}
0 & 1 & 0 \\
0 & 0 & 1 \\
\gamma^* & 0 & 0
\end{bmatrix}
=
\begin{bmatrix}
|\gamma|^2 & 0 & 0 \\
0 & 1 & 0 \\
0 & 0 & 1
\end{bmatrix}.
\]

The same computation holds for the last set of basis elements, involving the generator $\Gamma^2$.
\begin{eqnarray*}
&&\mu_1V_j\Gamma^2\mu_2^*\Gamma^{2*}V_k^*+\mu_2V_k\Gamma^2\mu_1^*\Gamma^{2*}V_j^* = 0
\end{eqnarray*}

At the price of sacrificing full-diversity, we may choose $\theta = -1$. This will result in additional relations: 
$$\alpha_\theta(D_i, 0)\alpha_\theta(0, D_j)^*+\alpha_\theta(0,D_j)\alpha_\theta(D_j, 0)^* = 0$$ for all $i,j = 1,  \ldots, 6$,
implying decoding complexity $O(|S|^{30})$ using \cite[Lemma 2]{JR}, { {rather than the worst case  $O(|S|^{36})$ for the given code. }}

{ {
\begin{example}
$\Dc = (L/K, \sigma_2, 3)$  with 
$L = \Q(\zeta_7)$, $\sigma_2:\zeta_7 \mapsto \zeta_7^2$
and $K = \Q(\sqrt{-7})$. 
To define the map $\alpha_\theta$,  we use $\tau: \sqrt{-7} \mapsto -\sqrt{-7}$ and 
consider $\theta = \sqrt{-7}$. 
\label{ex:6x3Example2}
\end{example}
}}

\begin{figure}
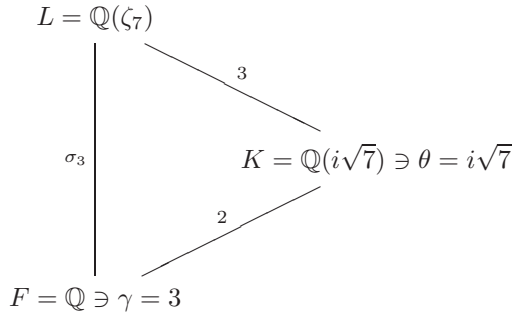

\[
\begin{diagram}
\node{L=\Q(\zeta_7)}\arrow{se,l,-}{3}\arrow[2]{s,l,-}{\sigma_3}\\
\node{}\node{K=\Q(i\sqrt{7})\ni\theta=i\sqrt{7}}\arrow{sw,l,-}{2}\\
\node{F=\Q\ni\gamma=3}
\end{diagram}
\]
\caption{
\label{fig:extcode2}
Field extension used for the second code example.
}
\end{figure}

\begin{claim}The algebra $\Dc = (L/K, \sigma_2, 3)$ is division.
\end{claim}
To see this, we use similar arguments as in Example \ref{ex:6x3Example1} to demonstrate that $3$ is a non-norm element in the extension $L/K$. Namely, $3$ multiplicatively generates nonzero elements of $\mathbb F_7$, which is the residue field of the ideal above $7$ in $K$. Therefore the image of $3$ in $\mathbb F_7^\times$ cannot lie in the kernel of the epimorphism $\F^\times_7 \rightarrow C_3$. 
Hence the algebra $\Dc = (L/K, \sigma_2, 3)$ is division.
The { {representation $\lambda$}} of $\Dc$ gives a family of $3\times3$ matrices of the form
{ {
$$\sum_{k=0}^2
\sum_{j=1}^3
c_{kj}
\begin{bmatrix}
\nu_j & 0 & 0 \\
0   & \sigma(\nu_j) & 0 \\
0 & 0 & \sigma^2(\nu_j)
\end{bmatrix}
\Gamma^k.$$
}}

where $c_{kj}  \in K$ and $\{\nu_1,\nu_2, \nu_3 \}$ is a $K$-basis of $L$. Each matrix encodes $9$ elements of $K$, or $18$ elements of $\Q$, since this time $K$ is of degree 2 over $\Q$. 
For the iterative construction, first we observe that $\theta=i\sqrt{7}$ is an element whose cube is not fixed by $\tau$. Hence, by Lemma \ref{ConditionFD3} the elements in the image of $\alpha_{i\sqrt{7}}$ have nonzero determinant, moreover since 
{ {
$K^{\<\tau\>} = \mathbb Q$,   Lemma \ref{normConditionN} implies that the resulting code has NVD. 
}}
We set the code to be the matrices in the image of $ \tilde \alpha_{i\sqrt{\theta}}(\Dc,\Dc)$, and by Lemma \ref{lem:fd},  fast-decodable properties of $\Dc$ are inherited. Similar computations as in the previous example can be done.

%***********************************************************************************************************************%
%
% CONCLUSION
%
%***********************************************************************************************************************%

\section{Conclusion and Future Work}

Motivated by the code design of fast decodable space-time codes for asymmetric MIMO channels, 
we proposed an iterated code construction, that maps a space-time block code coming from a cyclic division algebra into a new space-time code of twice the original dimension. 
We studied the algebraic structure of the newly obtained codes, which are actually forming interesting families of division algebras, whose center and maximal subfield are computed. 
In particular, we gave a condition for full diversity.
Particular attention was paid to the quaternion case, corresponding to the MIDO case.

It is surely possible to obtain other (possibly better) codes from the general construction that we proposed. It would be a valuable contribution to establish a bound on the best decoding complexity that codes coming from division algebras can have.
From a purely mathematical point of view, it would be of interest classify the division algebras that are built through the proposed iterative construction.

%*********************************************************************************************%
%
% ACK
%
%***********************************************************************************************************************%
\section*{Acknowledgments}

This research is supported by the Singapore National
Research Foundation under Research Grant NRF-CRP2-2007-03.

%%%%%%%%%%%%%%%%%%%%%%%%%%%%%%%%%%%%%%%%%%%%%%%%%%%%%%%%%%%%%%%%%%%%%%%%%%%%%%%%%%%%%%%%%%%%
%
% REFERENCES
%
%%%%%%%%%%%%%%%%%%%%%%%%%%%%%%%%%%%%%%%%%%%%%%%%%%%%%%%%%%%%%%%%%%%%%%%%%%%%%%%%%%%%%%%%%%%%
%\input{"bib.tex"}


\begin{thebibliography}{10}
%
\bibitem{VB} 
E. Viterbo, J. Boutros, ``A universal lattice decoder for fading channels,'' \emph{IEEE Trans. Inf. Theory}, vol. 45, no. 5, 1999.
%
\bibitem{BHV}
E. Biglieri, Y. Hong and E. Viterbo, ``On fast-decodable space-time block codes,''
{\em IEEE Trans. Inform. Theory}, vol. 55, no. 2, Feb 2009.
%
\bibitem{JR} 
G. R. Jithamitra, B. Sundar Rajan, 
``Minimizing the Complexity of Fast Sphere Decoding of STBCs,'' preprint, available at \url{http://arxiv.org/abs/1004.2844}
%
\bibitem{NR} L. P. Natarajan, B. S. Rajan, ``Fast group-decodable STBCs via codes over GF(4),'' 
\emph{Proc. IEEE Int. Symp. Inform. Theory}, Austin, TX, June 2010.
%
\bibitem{LS} Lakshmi Prasad Natarajan and B. Sundar Rajan,
``Fast-Group-Decodable STBCs via codes over GF(4): Further Results,''
{\em Proceedings of IEEE ICC 2011, (ICC'11)}, Kyoto, Japan, June 2011.
%
\bibitem{SBM}
M. Sinnokrot, J. R. Barry, and V. Madisetti, ``Embedded Alamouti Space-Time Codes for High Rate and Low Decoding Complexity,'' 
{\em Asilomar Conference on Signals, Systems, and Computers}, Pacific Grove, California, October 26-29, 2008.
%
\bibitem{Alamouti}
S.M. Alamouti, ``A simple transmit diversity technique for wireless communications,'' 
{\em IEEE Journal on Selected Areas in Communications}, vol. 16, no 8, October 1998.
%
\bibitem{SR} K. P. Srinath, B. S. Rajan, ``Low ML-decoding complexity, large coding gain, full-diversity STBCs for $2 \times 2$ and $4 \times 2$ MIMO systems,'' \emph{IEEE J. on Special Topics in Signal Processing: managing complexity in multi-user MIMO systems}, 2010
%
\bibitem{LO}
L. Luzzi, F. Oggier, ``A family of fast-decodable MIDO codes from crossed-product algebras over $\QQ$,''
\emph{Proc. IEEE Int. Symp. Inform. Theory}, St Petersburg, July 2011.
%
\bibitem{OVH}
R. Vehkalahti, C. Hollanti, F. Oggier, 
``Fast-Decodable Asymmetric Space-Time Codes from Division Algebras,'' 
{\em IEEE Transactions on Information Theory}, vol. 58, no. 4, April 2012.
%
\bibitem{NR2}
Lakshmi Prasad Natarajan and B. Sundar Rajan,
``Asymptotically-Optimal, Fast-Decodable, Full-Diversity STBCs,''
Proceedings of IEEE ICC 2011, (ICC'11), Kyoto, Japan, 06-09 June, 2011. 
%
\bibitem{Tarokh}
V. Tarokh, N. Seshadri, A. R. Calderbank, ``Space-time codes for high data rate wireless communication: performance criterion and code construction,'' 
{\em IEEE Trans. on Information Theory}, vol. 44, no. 2, March 1998.
%
\bibitem{BRV}
J.-C. Belfiore, G. Rekaya, and E. Viterbo, “The Golden Code: A 2×2 full rate Space-Time Code with Non Vanishing Determinants,” {\em IEEE Trans. on Inf. Theory}, vol. 51, no 4, April 2005.
%
\bibitem{HLRVV}
C. Hollanti, J. Lahtonen, K. Ranto, R. Vehkalahti, and E. Viterbo, ``On the Algebraic Structure of the Silver Code,''{\em IEEE Information Theory Workshop}, Portugal, 2008.
% \bibitem{HL}, C. Hollanti,  H.F. Lu ``Construction Methods for Asymmetric and Multiblock Space-Time Codes'', \emph{IEEE Trans Inf. Theory} 
%
\bibitem{SRS}
B. A. Sethuraman, B. S. Rajan, and V. Shashidhar, ``Full-diversity, high-rate space-time block codes from division
algebras,'' {\em IEEE Trans. on Information Theory}, vol. 49, no. 10, pp. 2596-2616, Oct. 2003.
%
\bibitem{mtns}
N. Markin and F. Oggier, ``Iterated Fast Decodable Space-Time Codes From Crossed-Products,'' Extended Abstract, {\em 20th International Symposium on Mathematical Theory of Networks and Systems}, Melbourne, Australia. 
%
\bibitem{MOisit}
N. Markin and F. Oggier, ``A Class of Iterated Fast Decodable Space-Time Codes for $2^n$ Tx Antennas,'' 
to appear in the proceedings of {\em ISIT 12}.
%


\bibitem{BO}
G. Berhuy and F. Oggier, ``Space-Time Codes from Crossed Product Algebras of Degree 4,''


\bibitem{UM} T. Unger, N. Markin, ``Quadratic Forms and Space-Time Block Codes from Generalized Quaternion and Biquaternion Algebras,'' {\em IEEE Transactions on Information Theory}, vol. 57, no. 9, Sept. 2011, availabe online at \url{http://arxiv.org/abs/0807.0199}.
%
\bibitem{LMG} J. Lahtonen, N. Markin, G. McGuire ``Construction of Space-Time Codes from Division Algebras with Roots of Unity as Non-Norm Elements'', \emph{IEEE Transactions on Information Theory}, vol. 54, no. 11, pp. 5231--5235, 2008.
%
\bibitem{HT}
A. Hottinen and O. Tirkkonen, ``Precoder designs for high rate space-time block codes,''  {\em Proc. Conference on Information Sciences and Systems}, 2004.
%
\bibitem{PGA}
J. Paredes, A.B. Gershman, and M. G. Alkhanari, ``A 2x2 space--time code with non-vanishing determinants and fast maximum likelihood decoding,'' {\em IEEE International Conference on Acoustics, Speech, and Signal Processing (ICASSP2007)}, Hawaii, USA, 2007.
%
\bibitem{OHV}
F. Oggier, C. Hollanti, and R. Vehkalahti,`` An Algebraic MIDO-MISO Code Construction," {\em IEEE International Conference on Signal Processing and Communications (SPCOM)}, India, 2010.
%
\bibitem{DV} P. Dayal and M. K. Varanasi, “An Optimal Two Transmit Antenna Space-Time Code And Its Stacked Extensions,” {\em IEEE Trans. Inf. Theory}, vol. 51, n. 12, pp. 4348-4355, Dec. 2005.
%
\bibitem{Jafarkhani}
H. Jafarkhani, ``A Quasi-Orthogonal Space-Time Block Code", {\em IEEE Transactions on
Communications}, vol. 49, no. 1, January 2001.
%
\bibitem{MO11}
N.~Markin, F.~Oggier, `` Iterated MIDO Space-Time Code Constructions,'' {\em Allerton Conference}, 2011.
%
\bibitem{MOspcom}
N.~Markin, F. ~Oggier, ``Fast Decodable Codes for $6$Tx-$3$Rx MIMO Systems'', {\em IEEE International Conference on Signal Processing and Communications (SPCOM)}, India, 2012.
\end{thebibliography}
\end{document}